\documentclass[11pt]{article}

\usepackage{mathrsfs,amsmath,amsbsy,amssymb,mcite,empheq}

\def\be{\begin{equation}}
\def\ee{\end{equation}}
\def\bea{\begin{eqnarray}}
\def\eea{\end{eqnarray}}

\DeclareTextFontCommand{\textwasy}{\wasyfamily}
\def \wasyfamily{\fontencoding{U}\fontfamily{wasy}\selectfont}
\def \thorn{{\wasyfamily\char105}}
\DeclareTextCommand{\dh}{OT1}{{\wasyfamily\char107}}
\newcommand{\tho}{{\textrm\thorn}}
\renewcommand{\eth}{{\textrm{\dh}}}

\newcommand{\mbar}{{\bar{m}}}

\renewcommand{\d}{\mathrm{d}}

\title{Uniqueness of the Kerr-de Sitter spacetime as an algebraically special solution in five dimensions} 

\author{Gabriel Bernardi de Freitas, Mahdi Godazgar and Harvey S. Reall\\ {\footnotesize Department of Applied Mathematics and Theoretical Physics, University of Cambridge}\\ {\footnotesize Wilberforce Road, Cambridge CB3 0WA, UK}\\ {\footnotesize G.B.Freitas@damtp.cam.ac.uk, M.M.Godazgar@damtp.cam.ac.uk, H.S.Reall@damtp.cam.ac.uk}}

\date\today

\begin{document}

\maketitle

\begin{abstract}
We determine the most general solution of the five-dimensional vacuum Einstein equation, allowing for a cosmological constant, with (i) a Weyl tensor that is type II or more special in the classification of Coley {\it et al}, (ii) a non-degenerate ``optical matrix'' encoding the expansion, rotation and shear of the aligned null direction. The solution is specified by three parameters. It is locally isometric to the 5d Kerr-de Sitter solution, or related to this solution by analytic continuation or taking a limit. This is in contrast with four dimensions, where there exist infinitely many solutions with properties (i) and (ii).
\end{abstract}

\section{Introduction}

There is a long tradition of classifying solutions of the four-dimensional Einstein equation according to the algebraic type of the Weyl tensor. If the Weyl tensor is everywhere type II, or more special, then the solution is called algebraically special. In an algebraically special spacetime, the Einstein equation simplifies considerably and one can determine the explicit dependence of the metric on one of the coordinates. The Einstein equation then reduces to PDEs in 3 dimensions. The general solution of these PDEs is not known. However, it is clear from various special cases that the general solution involves arbitrary functions \cite{exactsolutions}.

If one makes the stronger assumption that the Weyl tensor is (everywhere) of type D then much more progress can be made. The vacuum Einstein equation can be solved explicitly \cite{kinnersley}. There are several families of solutions, each specified by a few parameters. These solutions include the Kerr solution. Performing this classification led to the discovery of a new vacuum solution: the spinning C-metric, describing a pair of rotating black holes being accelerated by cosmic strings.

In this paper we will consider higher-dimensional solutions of the vacuum Einstein equation, allowing for a cosmological constant:
\be
\label{einstein}
 R_{ab} = \Lambda\, g_{ab}.
\ee
The algebraic classification of the Weyl tensor has been extended to $d$ spacetime dimensions by Coley {\it et al.} \cite{cmpp}. A solution is type II or more special in this classification if it admits a {\it multiple Weyl aligned null direction} (multiple WAND) $\ell$. The condition for $\ell$ to be a multiple WAND can be written \cite{ortaggio}
\be
 \ell^b  \ell_{[e} C_{a]b[cd} \ell_{f]} = 0,
\ee
where $C_{abcd}$ is the Weyl tensor. 
In four dimensions, this is the same as the condition for $\ell$ to be a repeated principal null direction. The Myers-Perry black hole solution \cite{myersperry} (``higher dimensional Kerr") is known to have a Weyl tensor of type D \cite{frolov,hourityped,typed}, which means that it admits two distinct multiple WANDs. The Myers-Perry solution has been generalized to include a cosmological constant in five \cite{hht} and higher \cite{glpp} dimensions. For simplicity, we will refer to these as ``Kerr-de Sitter" solutions (for any value of $\Lambda$). These also have a Weyl tensor of type D \cite{hourityped}. 

By analogy with the four-dimensional case, one might expect there to exist a small number of families of solutions of (\ref{einstein}) with a Weyl tensor of type D. Surprisingly, we will establish a stronger result in five dimensions: subject to one extra assumption, the Kerr-de Sitter solution is essentially the only solution with a Weyl tensor of type II.

To explain the extra assumption, we note that any solution of (\ref{einstein}) admitting a multiple WAND must also admit a geodesic multiple WAND \cite{geodesic} hence there is no loss of generality in assuming $\ell$ to be geodesic. Recall that the expansion, rotation and shear of the null geodesic congruence tangent to $\ell$ are defined as the trace, antisymmetric part, and traceless symmetric part of the $(d-2) \times (d-2)$ ``optical matrix''
\be
 \rho_{ij} = m_i^a m_j^b \nabla_b \ell_a ,
\ee 
where $m_i$ are a set of $(d-2)$ orthonormal spacelike vectors orthogonal to $\ell$. Our assumption is that $\rho_{ij}$ is non-degenerate.\footnote{Note that this does not depend on how the $m_i$ are chosen.} We will prove:

\medskip

\noindent {\bf Theorem.} Let $(M,g)$ be a solution of the five-dimensional vacuum Einstein equation (\ref{einstein}). Assume that $(M,g)$ admits a geodesic multiple WAND $\ell$ for which the $3\times 3$ matrix $\rho_{ij}$ is non-degenerate. Then one can define an affine parameter $r$ along the null geodesics tangent to $\ell$ such that the eigenvalues of $\rho_{ij}$ are $1/r$ and $1/(r \pm i \chi)$ for some real function $\chi$ constant along each geodesic. Furthermore:

\begin{enumerate}
\item
If $\chi \ne 0$ and $d\chi \ne 0$ then one can define local coordinates $(u,r,\chi,x,y)$ such that the metric is
\bea
\label{unequal}
 ds^2 &=& - 2 (\d u + \chi^2 \d y) \left[ \d r + H(r,\chi) (\d u + \chi^2 \d y) - E_0 \left( \d x - \frac{E_0}{\chi^2} \d y \right) + P \d y \right] \nonumber \\
      &\qquad& + r^2 \chi^2 \left( \d x - \frac{E_0}{\chi^2} \d y \right)^2 + (r^2 + \chi^2)  \left( \frac{\d \chi^2}{P(\chi)} + P(\chi) \d y^2 \right),
\eea
where
\be
\label{Hdef}
H(r,\chi) = A_0 - \frac{\Lambda}{8} (r^2 - \chi^2) - \frac{\mu_0}{2(r^2 + \chi^2)},
\ee
\be
 P(\chi) =  C_0 - \frac{E_0^2}{\chi^2} - 2 A_0 \chi^2 - \frac{\Lambda \chi^4}{4} 
\ee
$(A_0,\mu_0,C_0,E_0)$ are arbitrary real constants, and $P(\chi)>0$. 
\item
If $\chi \ne 0$ and $d\chi \equiv 0$ then one can define local coordinates $(u,r,x,y^1,y^2)$ such that the metric is
\bea
\label{equal}
 ds^2 &=& - 2 (\d u + 2 \chi \mathcal{A}) \left[ \d r + H(r) (\d u + 2 \chi \mathcal{A} ) - E_0 \left( \d x + \frac{2 E_0}{\chi^3} \mathcal{A} \right) \right] \nonumber \\
      &\qquad& + r^2 \chi^2 \left( \d x + \frac{2 E_0}{\chi^3} \mathcal{A} \right)^2 + (r^2 + \chi^2) h_{\alpha \beta}(y) dy^\alpha dy^\beta,
\eea
where $h_{\alpha \beta}(y)$ ($1\le \alpha,\beta \le 2$) is the metric on a 2d Riemannian manifold of constant curvature, $\mathcal{A} = \mathcal{A}_\alpha(y) dy^\alpha$ is a 1-form such that $d \mathcal{A}$ is a volume form for this 2d manifold,
\be
 H(r ) =  \frac{E_0^2}{2\chi^4}  - \frac{\Lambda}{8} (r^2 + \chi^2) - \frac{\mu_0}{2(r^2 + \chi^2)}
\ee
and $(\chi,\mu_0,E_0)$ are arbitrary real constants (with $\chi \ne 0$). The Ricci scalar of $h_{\alpha\beta}$ is
\be
 R^{(2)} = \frac{8 E_0^2}{\chi^4} + 2 \Lambda \chi^2.
\label{const:curv}
\ee
\item
If $\chi \equiv 0$ then one can define local coordinates $(u,r,y^1,y^2,y^3)$ such that the metric is
\be
\label{schw}
 ds^2 = -\left( k - \frac{\mu_0}{r^2} -\frac{\Lambda}{4} r^2 \right) du^2 - 2 du dr + r^2 h_{ij} (y) dy^i dy^j,
\ee
where $k \in \{-1,0,1\}$, $\mu_0$ is a constant, and $h_{ij}(y)$ ($1 \le i,j \le 3)$ is the metric on a 3d Riemannian space of constant curvature with Ricci scalar $6k$. 
\end{enumerate}

\medskip

We will now make some remarks on the above theorem. 

For any of the above metrics $\partial/\partial r$ is a geodesic multiple WAND. The Weyl tensor vanishes if, and only if, $\mu_0 =0$. For $\mu_0 \ne 0$ each solution is type D, i.e., there exists a second multiple WAND. 

The metric (\ref{unequal}), has a scaling symmetry: for $\lambda \ne 0$ one can perform a coordinate transformation
\be
u = \frac{u'}{\lambda} \qquad r = \lambda r' \qquad \chi = \lambda \chi' \qquad x = \frac{x'}{\lambda^2} \qquad y = \frac{y'}{\lambda^3}
\ee 
and the metric in the primed coordinates takes the same form as (\ref{unequal}) but with rescaled constants:
\be
\label{scaling}
 A_0' = \frac{A_0}{ \lambda^2} \qquad \mu_0' = \frac{\mu_0}{\lambda^4} \qquad C_0' = \frac{C_0}{\lambda^4} \qquad E_0' = \frac{E_0}{\lambda^3}.
\ee 
This shows that the solution is really a 3-parameter family.\footnote{We choose not to use this symmetry to eliminate one of the constants because this requires consideration of various special cases e.g. if $E_0 \ne 0$ then one can rescale to set $E_0=1$ but then the case $E_0=0$ needs to be discussed separately.} 

We will show (in section \ref{sec:newcoords}) that the solution (\ref{unequal}) is locally isometric to the 5d Kerr-de Sitter solution \cite{hht} with two unequal rotation parameters.\footnote{Except in the special case $\Lambda=A_0 = E_0 =0$, $C_0 = \mu_0$, which can only be obtained as a limit of the Kerr-de Sitter solution. This is the only non-trivial case for which the second multiple WAND has degenerate (in fact vanishing) optical matrix.} The coordinates of (\ref{unequal}) are closely related to the coordinates for the Kerr-de Sitter solution defined in Ref.~\cite{chen}. Only when the parameters $(A_0,\mu_0,C_0,E_0)$ lie within a certain set does the metric (\ref{unequal}) describe a regular black hole solution. Values of the parameters outside this set can result in local metrics such as the ``Kaluza-Klein bubble'' spacetime of Ref.~\cite{bubble} which was shown to be algebraically special in Ref.~\cite{gspaper}.\footnote{This spacetime was originally obtained by analytic continuation of the "Boyer-Lindquist" coordinates of the Myers-Perry solution. No analytic continuation of the coordinates is required to obtain it in the coordinates of (\ref{unequal}).}

The metric (\ref{equal}) has a scaling symmetry analogous to (\ref{scaling}). Hence this solution is really a 2-parameter family. If $R^{(2)}>0$, one can perform a coordinate transformation to show that the solution is locally isometric to the Kerr-de Sitter solution with two equal, non-zero, rotation parameters. Solutions with $R^{(2)} \le 0$ can be regarded as analytically continued versions of the Kerr-de Sitter solution. 

The metric (\ref{schw}) is a generalized Schwarzschild metric written in outgoing Eddington-Finkelstein coordinates. Of course, this corresponds to the Kerr-de Sitter metric with vanishing rotation parameters.

Note that, in four dimensions, there are many solutions satisfying the assumptions of the above theorem. In 4d, the Goldberg-Sachs theorem implies that $\ell$ is shear-free, which implies that $\rho_{ij}$ is degenerate if, and only if, $\rho_{ij} = 0$ (this defines the Kundt family of solutions). So any algebraically special solution with $\rho_{ij} \ne 0$ satisfies the assumptions of the theorem. As noted above, there is no simple explicit form for such solutions, and such solutions are known to involve free functions. 

The non-degeneracy assumption on $\rho_{ij}$ cannot be eliminated from our theorem. To see this, note that one can take the product of a 4d Ricci flat algebraically special solution with a flat direction to obtain a 5d Ricci flat solution admitting a geodesic multiple WAND. This solution will have degenerate $\rho_{ij}$ because $\ell$ does not expand along the flat direction. Obviously there are as many such solutions as there are 4d algebraically special solutions. For $\Lambda \ne 0$ one can take a warped product to reach the same conclusion.

Our result can be viewed as a new kind of uniqueness theorem for the 5d Kerr-de Sitter solution. It should be contrasted with the usual uniqueness theorem \cite{morisawa,hollands} for the 5d Myers-Perry black hole, which assumes the existence of a $\mathbb{R} \times U(1)^2$ isometry group, asymptotic flatness, a regular horizon of spherical topology, and no topology outside the horizon. Our result assumes nothing about isometries or global structure and allows for a cosmological constant. 

The local nature of our result is similar to the uniqueness result for spacetimes with certain ``hidden symmetries'', i.e., symmetries associated to Killing tensors rather than Killing vectors. Refs. \cite{houri,krtous} proved that the Kerr-de Sitter solution (generalized to allow for a NUT charge \cite{chen}) is the most general $d$-dimensional solution of (\ref{einstein}) admitting a ``principal conformal Killing-Yano 2-form''. Note that this result applies even for $d=4$, where it yields a subset of the type D solutions. This is in contrast with our result, for which there exist infinitely many $d=4$ solutions satisfying the assumptions of the theorem. 

Our result is also reminscent of the theorem that asserts that the Kerr solution is the unique stationary solution of the 4d vacuum Einstein equation with vanishing Mars-Simon tensor \cite{mars}. However our theorem does not assume stationarity. 

There have been several hints that higher-dimensional algebraically special solutions with non-degenerate $\rho_{ij}$ might be more rigid than their 4d counterparts. First, Ref.~\cite{hidrt} considered higher dimensional Robinson-Trautman solutions, defined by the existence of a null geodesic $\ell$ with $0 \ne \rho_{ij} \propto \delta_{ij}$ (such $\ell$ must be a multiple WAND). It was found that these solutions are considerably simpler for $d>4$ than for $d=4$. For $d=5$, the only such solution is the generalized Schwarzschild metric (\ref{schw}). 

Second, Ref.~\cite{dias} investigated the possible existence of families of algebraically special solutions that contain the Schwarzschild solution. A solution ``close'' to the Schwarzschild solution in such a family would have non-degenerate $\rho_{ij}$. The approach of Ref.~\cite{dias} was to consider linear perturbations of the Schwarzschild solution that preserve the algebraically special property. For $d=4$ there are infinite families of such perturbations. But for $d>4$ it was found that the only such perturbations that are regular on the orbits of spherical symmetry are perturbations corresponding to a linearisation of the Myers-Perry solution around the Schwarzschild solution. 

Third, Ref.~\cite{asympflat} studied algebraically special solutions in $d>4$ dimensions that (i) have non-degenerate $\rho_{ij}$ and (ii) are asymptotically flat. It was assumed that the curvature components can be expanded in inverse powers of an affine parameter $r$ along the null geodesics tangent to $\ell$. It was found that such solutions are non-radiative, in contrast with the $d=4$ case. 

Ref.~\cite{gspaper} showed that 5d algebraically special solutions can be classified according to the rank of $\rho_{ij}$. Our theorem determines all solutions for which $\rho_{ij}$ has rank 3. Rank 0 defines the Kundt class of solutions  which was studied in Refs \cite{kundt1,kundt2}. For rank 2 or rank 1, all solutions for which $\ell$ is {\it hypersurface-orthogonal} ($\rho_{[ij]}=0$) were determined in Ref.~\cite{hyporthog}. We intend to return to the general (non-hypersurface-orthogonal) rank 2 and rank 1 cases in future work.  

We end this introduction with an outline of the proof of our theorem. 
The starting point for the proof is the recent demonstration \cite{gspaper} that the optical matrix of a geodesic multiple WAND in 5d can be brought to a certain canonical form by an appropriate choice of the basis vectors $m_i$. This is the 5d analogue of the ``shearfree'' property that holds in 4d because of the Goldberg-Sachs theorem. As noted in Ref.~\cite{gspaper}, case 3 of the theorem follows immediately from combining this canonical form with the results of Ref.~\cite{hidrt}. 

For non-degenerate $\rho_{ij}$, the canonical form involves two unknown functions. We show how the evolution equation for $\rho_{ij}$ can be integrated to determine the dependence of these functions on an affine parameter $r$ along the geodesics tangent to $\ell$. This determines the form of the eigenvalues of $\rho_{ij}$ as stated in the theorem.

Next we complete $\{\ell,m_i\}$ to a null basis $\{\ell,n,m_i\}$ where $n$ is null and orthogonal to $m_i$. After exploiting a residual freedom in the choice of $m_i$, we show how the ``Newman-Penrose'' and Bianchi equations can be integrated to determine the $r$-dependence of the basis vectors and hence the $r$-dependence of the metric. The $r$-dependence of the connection and curvature components is also fully determined. This calculation reveals that the Weyl tensor is necessarily of type D. 

The vanishing of certain connection components enables us to introduce local coordinates in a canonical way. After expressing our basis vectors in terms of these coordinates and using the results obtained previously we obtain a set of equations that can be integrated. At this stage, it becomes convenient to divide the analysis into two cases depending on whether $\chi$ is constant or not. If $d\chi \ne 0$ then we use $\chi$ as a coordinate and show that residual coordinate freedom can be exploited to make the solution independent of three of the remaining coordinates. Finally we solve for the dependence on $\chi$ to obtain the solution (\ref{unequal}). If $d\chi \equiv 0$ then a similar procedure leads to the metric (\ref{equal}). 

This paper is organized as follows: section \ref{sec:basis} contains the first part of the proof of the theorem in which we determine the connection and curvature components in a null basis. In section \ref{sec:coords}, we introduce coordinates and complete the proof of the theorem. Section \ref{sec:newcoords} demonstrates how the metrics (\ref{unequal}) and (\ref{equal}) are related to the Kerr-de Sitter solution. 

\subsection*{Notation}

We will perform most calculations in a null basis. Refs. \cite{bianchi,ricci} developed a higher-dimensional analogue of the Newman-Penrose formalism used for calculations in such a basis. This was repackaged into a  higher-dimensional analogue of the Geroch-Held-Penrose (GHP) formalism in Ref.~\cite{ghp}. We will follow the notation of Ref.~\cite{ghp} for the connection components and Weyl tensor components. In particular, we refer the reader to eqns.~NP1--NP4, B1--B8 and C1--C3 of Ref.~\cite{ghp}, which lists all the Newman-Penrose and Bianchi equations satisfied by the connection and curvature components, as well as equations for the commutator of derivatives.

\section{Integration of GHP equations}

\label{sec:basis}

\subsection{Canonical form for $\rho_{ij}$}

Consider an Einstein spacetime, i.e. a solution of \eqref{einstein}, admitting a multiple WAND $\ell$. Introduce a null basis $\{\ell,n,m_i\}$, $i=2,3,4$. The multiple WAND condition is equivalent to the vanishing of the Weyl components of boost weights $+2$ and $+1$:
\be
\Omega_{ij} = 0, \qquad \Psi_{ijk} = 0.
\ee
Without loss of generality, $\ell$ can be assumed to be geodesic \cite{geodesic}, i.e. 
\be
 \kappa_i = 0.
\ee
Ref.~\cite{gspaper} showed that the spatial basis vectors $m_i^a$ ($i=2,3,4$) can be chosen so that the optical matrix $\rho_{ij}$ of $\ell$ takes one of the following forms:
\be \label{rho}
 b\left( \begin{array}{lll} 1 & a & 0 \\ -a & 1 & 0 \\ 0 & 0 & 1+a^2 \end{array} \right) \qquad b\left( \begin{array}{lll} 1 & a & 0 \\ -a & 1 & 0 \\ 0 & 0 & 0 \end{array} \right)\ \qquad b\left( \begin{array}{lll} 1 & a & 0 \\ -a & -a^2 & 0 \\ 0 & 0 & 0 \end{array} \right)
\ee
for functions $a,b$. For $b \ne 0$, these matrices have rank 3,2,1 respectively. Our assumption is that $\rho_{ij}$ is non-degenerate so
\be
 \rho_{ij} = b \left(
               \begin{array}{lll}
                1  & a & 0 \\
                -a & 1 & 0 \\
                0  & 0 & 1 + a^2
               \end{array}
               \right), \qquad b \ne 0.
\ee
We know all the GHP scalars with positive boost weight, both from the connection and the curvature: $\Omega_{ij}, \Psi_{ijk}, \kappa_i$, and we know the structure of $\rho_{ij}$, which has boost weight $+1$. Using the Newman-Penrose equations and Bianchi identities, we can determine the GHP scalars of boost weight negative and zero by systematically examining these equations from higher to lower boost weight. We now proceed to indicate the steps involved in this calculation.

\subsection{Choice of basis}

The form of the optical matrix above does not fix the basis uniquely because this form is preserved by null rotations about $\ell$ and spins in the 2-3 directions. We can use this freedom to make some of the GHP scalars vanish. 
Consider a null rotation about $\ell$ with parameters $z_i$ \cite{ghp}
\be
 \ell \mapsto \ell, \qquad n \mapsto n + z_i m_i - \frac{1}{2} z^2 \ell, \qquad m_i \mapsto m_i - z_i \ell,
\label{null:rotation}
\ee
where $z^2 = z_i z_i$. This leaves $\rho_{ij}$ unchanged (as $\kappa_i=0$) but $\tau_i$ changes according to \cite{ghp}
\be
 \tau_i \mapsto \tau_i + \rho_{ij} z_j.
\ee
Since $\rho_{ij}$ is non-degenerate, we can choose $z_i$ to set $\tau_i = 0$. Equation NP2 of Ref.~\cite{ghp} then gives $\tau'_i = 0$ \cite{ghp}. In summary, we choose our basis so that:
\be
 \tau_i = \tau'_i = 0.
\ee
It is convenient to combine the spatial vectors $m_2,m_3$ into complex null vectors:
\be
 m_5 = \frac{m_2 + i m_3}{\sqrt{2}}, \qquad \mbar_5 = \frac{m_2 - i m_3}{\sqrt{2}}.
\ee
In this frame, we have\footnote{
We will use indices $i,j,\ldots$ to label both the real basis ($i,j=2,3,4$) and the new basis $(i,j=4,5,\bar{5})$. The meaning should be clear from the context.}
\be
 \delta_{ij} = \left(
    \begin{array}{ccc}
      1 & 0 & 0 \\
      0 & 0 & 1 \\
      0 & 1 & 0
    \end{array}
               \right),
\ee
while the optical matrix is written as
\be
 \rho_{ij} = b \left(
    \begin{array}{ccc}
      1 + a^2 & 0       & 0 \\
      0       & 0       & 1 - i a \\
      0       & 1 + i a & 0
    \end{array}
                \right).
\ee
Now consider a spin in the 2-3 directions, which can be phrased in terms of the null complex frame as
\be
 m_5 \mapsto e^{i \lambda} m_5, \qquad \mbar_5 \mapsto e^{- i \lambda} \mbar_5,
\label{spin}
\ee
for some function $\lambda$. This will induce changes in $\stackrel{i}{M}_{j0}$ as follows (recall that $D \equiv \ell \cdot \partial$):
\bea
 \stackrel{4}{M}_{50}        &\mapsto& e^{i \lambda} \stackrel{4}{M}_{50}, \\
 \stackrel{5}{M}_{\bar{5} 0} &\mapsto& \stackrel{5}{M}_{\bar{5} 0} + i D\lambda.
\eea
The last equation, in particular, implies that we can choose $\lambda$ to set $\stackrel{5}{M}_{\bar{5} 0} = 0$ (note that the LHS is imaginary). Moreover, eqn.~NP1 of Ref.~\cite{ghp} for $ij = 45$ gives $\stackrel{4}{M}_{50} = 0$ which, from the above, is preserved under such spins. Hence we have 
\be
\stackrel{i}{M}_{j0} = 0. 
\ee
Finally, we are free to rescale $\ell$ so that the geodesics with tangent $\ell$ are affinely parameterized, which implies $L_{10}=0$. Note that the conditions
\be
 L_{10}=0, \qquad \kappa_i = 0, \qquad \tau'_i = 0, \qquad \stackrel{i}{M}_{j0} = 0,
\ee
mean that we have chosen our basis to be parallelly transported along the geodesics with tangent $\ell$.

\subsection{Determining $a,b$}

We introduce local coordinates as follows. Pick a hypersurface $\Sigma$ transverse to $\ell$ and introduce coordinates $x^\mu$ on $\Sigma$. Now assign coordinates $(r,x^\mu)$ to the point parameter distance $r$ along the integral curve of $\ell$ through the point on $\Sigma$ with coordinates $x^\mu$. Note that $r$ is an affine parameter along the geodesics. We now have
\be
 \ell = \frac{\partial}{\partial r}.
\ee
Consider eqn.~NP1 of Ref.~\cite{ghp}, which in our parallelly transported basis reads
\be
 D \rho_{ij} = - \rho_{ik} \rho_{kj}.
\label{A2}
\ee
Taking the $ij = 5\bar{5}$ component gives:
\be
 \frac{\partial}{\partial r} \left[ b(1 - ia) \right] = - \left[ b (1 - ia) \right]^2.
\label{Da-Db}
\ee
The solution is then
\be
 b (1 - ia) = \frac{1}{r - i \chi},
\ee
for some complex function $\chi$ that does not depend on $r$. There is freedom in defining $r$ in the sense that we can shift it by a function of the other coordinates, $r \rightarrow r + \alpha(x^{\mu})$, which corresponds to moving the surface $\Sigma$ used to define $r$. This freedom can be used to set $\mathrm{Im} (\chi) = 0$. With $\chi$ real, we take the real and imaginary parts of the above equation to find
\be
 a = - \frac{\chi}{r}, \qquad b = \frac{r}{r^2 + \chi^2}.
\label{a-b:r}
\ee
The $r$-dependence of $\rho_{ij}$ is then given by
\be
 \rho_{ij} = \left(
    \begin{array}{ccc}
      \frac{1}{r} & 0                    & 0 \\
      0           & 0                    & \frac{1}{r - i \chi} \\
      0           & \frac{1}{r + i \chi} & 0
    \end{array}
                \right).
\ee
In the real basis, the eigenvalues of $\rho_{ij}$ are $b(1+a^2)$ and $b(1 \pm i a)$. Using (\ref{a-b:r}) we see that these are $1/r$ and $1/(r \pm i \chi)$, as asserted in our Theorem. 

\subsection{The case $\chi \equiv 0$}

If $\chi \equiv 0$ then we have $\rho_{ij} = (1/r) \delta_{ij}$, so $\ell$ is free of rotation and shear with non-vanishing expansion. This defines the Robinson-Trautman class of solutions. These solutions were studied in Ref.~\cite{hidrt}, where is was proved that the only solution of this type (in 5d) is the generalized Schwarzschild solution (\ref{schw}). This establishes case 3 of the Theorem. Henceforth we will assume $\chi \ne 0$. 

\subsection{Boost weight 0 components of the Weyl tensor}

The only GHP scalars with boost weight 0 that we do not know yet are such components of the Weyl tensor: $\Phi_{ijkl}, \Phi_{ij}$. Their $r$- dependence can be determined completely using the Bianchi identities, as we will now explain. Notice that, in five dimensions, all the information regarding the boost weight 0 components of the Weyl tensor is encoded in $\Phi_{ij}$, for one can write 
\be
 \Phi_{ijkl} = -2 \left( \Phi^{\mathrm{S}}_{ik} \delta_{jl} - \Phi^{\mathrm{S}}_{il} \delta_{jk} - \Phi^{\mathrm{S}}_{jk} \delta_{il} + \Phi^{\mathrm{S}}_{jl} \delta_{ik} \right) + \Phi \left( \delta_{ik} \delta_{jl} - \delta_{il} \delta_{jk} \right)
\ee
in terms of the symmetric part $\Phi^{\mathrm{S}}_{ij} = \Phi_{(ij)}$.

The relevant Bianchi identities to determine $\Phi_{ij}$ are equations B2, B3 and B4 of Ref.~\cite{ghp}, which become, in our basis,
\be
 D \Phi_{ij} = - \left( \Phi_{ik} + 2 \Phi_{ik}^{\mathrm{A}} + \Phi \delta_{ik} \right) \rho_{kj},
\label{A10}
\ee
\be
 D \Phi_{ijkl} = - 2 \Phi^{\mathrm{A}}_{ij} (\rho_{kl} - \rho_{lk}) + \Phi_{ki} \rho_{jl} - \Phi_{li} \rho_{jk} - \Phi_{kj} \rho_{il} + \Phi_{lj} \rho_{ik} - \Phi_{ijkm} \rho_{ml} + \Phi_{ijlm} \rho_{mk} \label{A11}
\ee
and
\be
 0 = 2 \Phi^{\mathrm{A}}_{[jk|} \rho_{i|l]} - 2 \Phi_{i[j} \rho_{kl]} + \Phi_{im[jk|} \rho_{m|l]},
 \label{A12}
\ee
respectively.

Take first the $ijkl = 545\bar{5}$ component of \eqref{A12}, which gives
\be
 \Phi^{\mathrm{A}}_{45} = \frac{i \chi (2r + i \chi)}{r (r + 2 i \chi)} \Phi^{\mathrm{S}}_{45}.
\ee
Taking the $ijkl = 455\bar{5}$ component of \eqref{A11} and substituting the expression for $\Phi^{\mathrm{A}}_{45}$ then gives
\be
 2 D \Phi^{\mathrm{S}}_{45} + \left( \frac{5}{2r} + \frac{4}{r + i \chi} - \frac{3}{2r + 4 i \chi} \right) \Phi^{\mathrm{S}}_{45} = 0.
\ee
Comparing the previous two equations with the one obtained from the $ij = 54$ component of \eqref{A10} gives $\Phi^{\mathrm{S}}_{45} = 0$, and hence $\Phi_{45} = \Phi_{54} = 0$.

Now we compare the two equations that can be obtained for $\Phi_{55}$. The first comes from setting $ij = 55$ in \eqref{A10},
\be
 D \Phi_{55} + \frac{1}{r + i \chi} \Phi_{55} = 0.
\ee
The second is obtained by putting $ijkl = 4545$ in \eqref{A11},
\be
 2 D \Phi_{55} + \frac{5 r + 3 i \chi}{r (r + i \chi)} \Phi_{55} = 0.
\ee
One then immediately sees that consistency requires $\Phi_{55} = 0$.

Thus, the nontrivial components of $\Phi_{ij}$ can only be $\Phi_{44}, \Phi_{5\bar{5}}$. Setting $ijkl = 445\bar{5}$ in \eqref{A12} gives the relation
\be
 \Phi^{\mathrm{A}}_{5\bar{5}} = \frac{2 i \chi r}{r^2 + \chi^2} \Phi_{44}.
\ee
If we now substitute this into the $ijkl = 454\bar{5}$ component of \eqref{A11}, we find
\be
 D \Phi_{44} + \frac{3r^2 + \chi^2}{r(r^2 + \chi^2)} \Phi_{44} + \frac{1}{r} \Phi^{\mathrm{S}}_{5\bar{5}} = 0.
\ee
On the other hand, the $ij = 44$ component of \eqref{A10} gives
\be
 D \Phi_{44} + \frac{2}{r} \left( \Phi_{44} + \Phi^{\mathrm{S}}_{5\bar{5}} \right) = 0.
\label{A10:44}
\ee
Comparing these we find
\be
 \Phi^{\mathrm{S}}_{5\bar{5}} = \frac{r^2 - \chi^2}{r^2 + \chi^2} \Phi_{44}
\ee
and hence
\be
 \Phi_{5\bar{5}} = \frac{(r + i \chi)^2}{r^2 + \chi^2} \Phi_{44}.
\label{Phi55bar:Phi44}
\ee
Substituting this back into \eqref{A10:44}, we have
\be
 D \Phi_{44} = \frac{\partial \Phi_{44}}{\partial r} = - \frac{4r}{r^2 + \chi^2} \Phi_{44},
\ee
for which the solution is simply
\be
 \Phi_{44} = - \frac{\mu}{(r^2 + \chi^2)^2},
\ee
for some function $\mu$ independent of $r$. Therefore, the form of $\Phi_{ij}$ is
\be
 \Phi_{ij} = - \frac{\mu}{(r^2 + \chi^2)^3} \left(
                                          \begin{array}{ccc}
                                           r^2 + \chi^2 & 0 & 0 \\
                                           0 & 0 & (r + i \chi)^2 \\
                                           0 & (r - i \chi)^2 & 0
                                          \end{array}
                                        \right).
\label{Phi}
\ee
One can then verify that any other component of equations \eqref{A10}, \eqref{A11}, \eqref{A12} is trivially satisfied.

Now recall that we have chosen the normalization of $\ell$ such that it is tangent to affinely parameterized null geodesics. Choosing the affine parameter as one of the coordinates, as above, allows $\ell$ to be written as $\ell = \partial/\partial r$. This, however, does not determine $r$ uniquely. Previously we have used a shift in $r$ by the other coordinates to make $\chi$ real. Now consider the effect of a boost $\hat{\ell} = \lambda \ell$ for some non-zero function $\lambda$. The new vector field $\hat{\ell}$ will also be tangent to affinely parameterized geodesics provided that $\lambda$ is independent of $r$. Thus, we can define
\be
 \hat{\ell} = \frac{\partial}{\partial \hat{r}},
\ee
where $\hat{r} = r/\lambda$. If the analysis above were repeated using $\hat{r}$ instead of $r$, the optical matrix $\hat{\rho}_{ij}$ of $\hat{\ell}$ would have the same form as $\rho_{ij}$ if we defined $\hat{\chi} = \chi/\lambda$. Note that this is consistent with the fact that $\rho_{ij}$ transforms with boost weight $+1$, $\hat{\rho}_{ij} = \lambda \rho_{ij}$. On the other hand, $\Phi_{ij}$ has boost weight $0$ and so is invariant under boosts. Therefore it would retain the same form as in eqn.~\eqref{Phi} by defining $\hat{\mu} = \mu/\lambda^4$. But this shows that we can choose $\lambda$ so as to make $\hat{\mu}$ constant (but we can't choose its sign). Dropping the hats, we can assume, without any loss of generality, that 
\be
\mu = \mu_0,
\ee
where $\mu_0$ is a constant. Note that $\Phi_{ij}$ vanishes if, and only if, $\mu_0=0$.

\subsection{Optical matrix $\rho'_{ij}$ of $n^a$}

Having determined the $r$-dependence of $\Phi_{ij}$, the $r$-dependence of $\rho'_{ij}$ can now be completely determined by using information from eqns.~NP4, NP4$'$ of Ref.~\cite{ghp}. In our basis, these read ($\Delta \equiv n \cdot \partial$)
\be
 \Delta \rho_{ij} - L_{11} \rho_{ij} + \stackrel{k}{M}_{i1} \rho_{kj} + \stackrel{k}{M}_{j1} \rho_{ik} = - \rho_{ik} \rho'_{kj} - \Phi_{ij} - \frac{\Lambda}{4} \delta_{ij}
\label{A5}
\ee
and
\be
 D \rho'_{ij} = - \rho'_{ik} \rho_{kj} - \Phi_{ji} - \frac{\Lambda}{4} \delta_{ij},
\label{A6}
\ee
respectively.

We start by taking the $ij = 44$ component of \eqref{A6}, which gives
\be
 D \rho'_{44} = - \frac{\rho'_{44}}{r} + \frac{\mu_0}{(r^2 + \chi^2)^2} - \frac{\Lambda}{4}.
\ee
The solution to this equation is
\be
 \rho'_{44} = \frac{A}{r} - \frac{\mu_0}{2r(r^2 + \chi^2)} - \frac{\Lambda r}{8},
\ee
where $A$ is some function independent of $r$. Similarly, the $ij = 45$ component of the same equation yields
\be
 D \rho'_{45} = -\frac{\rho'_{45}}{r + i \chi},
\ee
to which the solution is
\be
 \rho'_{45} = \frac{B_5}{r + i \chi}
\ee
for some complex function $B_5$ that does not depend on $r$. Substituting this into \eqref{A5} for $ij = 45$, one finds
\be
 \stackrel{4}{M}_{51} = \frac{i B_5}{\chi}
\label{M451}
\ee
and then the $ij = 54$ component of \eqref{A5} determines
\be
 \rho'_{54} = - \frac{B_5}{r}.
\ee
In addition, \eqref{A5} with $ij = 55$ immediately gives
\be
 \rho'_{55} = 0.
\ee
Next, putting $ij = 5\bar{5}$ in eqn.~\eqref{A6} gives the equation
\be
 D \rho'_{5 \bar{5}} = - \frac{\rho'_{5 \bar{5}}}{r - i \chi} + \frac{\mu_0 (r - i \chi)^2}{(r^2 + \chi^2)^3} - \frac{\Lambda}{4}.
\ee
This can be integrated to give
\be
 \rho'_{5\bar{5}} = - \frac{\mu_0 (r - i \chi)}{2(r^2 + \chi^2)^2} + \frac{1}{r - i \chi} \left[ A_5 - \frac{\Lambda}{8} (r + i \chi) (r - 3 i \chi) \right],
\ee
where $A_5$ is a complex function independent of $r$. No further information can then be extracted from eqn.~\eqref{A6}.

Consider now the $ij = 44$ component of \eqref{A5},
\be
 n^r + r L_{11} = r \rho'_{44} - \frac{\mu_0 r^2}{(r^2 + \chi^2)^2} + \frac{\Lambda r^2}{4},
\label{A5:44}
\ee
where $n^r$ is the $r$-component of $n^a$ in the coordinate basis defined by $(r,x^\mu)$. Similarly the $ij = 5\bar{5}$ component gives
\be
 n^r - i \Delta \chi + (r - i \chi) L_{11} = (r - i \chi) \rho'_{5 \bar{5}} - \frac{\mu_0}{r^2 + \chi^2} + \frac{\Lambda}{4} (r - i \chi)^2,
\label{A5:55bar}
\ee
where we have used $\Delta r = n^r$. Subtracting \eqref{A5:44} from \eqref{A5:55bar} gives the relation
\be
 A_5 - A + i \Delta \chi + i \chi L_{11} + \frac{i \mu_0 r \chi}{(r^2 + \chi^2)^2} - \frac{\Lambda \chi}{8} (2 i r + 5 \chi) = 0.
\ee
Taking the real part yields simply
\be
 \mathrm{Re} (A_5) = A + \frac{5 \Lambda \chi^2}{8}
\ee
and hence one can write
\be
 \rho'_{5\bar{5}} = \frac{i \Lambda \chi}{4} - \frac{\mu_0 (r - i \chi)}{2 (r^2 + \chi^2)^2} + \frac{1}{r - i \chi} \left( A - i F - \frac{\Lambda r^2}{8} \right),
\ee
where we have redefined $\mathrm{Im} (A_5) = -F$. Thus the $r$-dependence of $\rho'_{ij}$ is now known. There is still some information left from eqns.~\eqref{A5:44} and \eqref{A5:55bar}, which will be exhausted in the next calculation.

\subsection{Determining the basis vectors}

In this section we determine the $r$-dependence of the basis vectors. It is sufficient to consider the commutators
\bea
\left[ \ell, n \right]   &=& - L_{11} \ell, \label{l:n:comm} \\
\left[ \ell, m_i \right] &=& - L_{1i} \ell - \rho_{ji} m_j \label{l:mi:comm},
\eea
together with the remaining information from eqn.~\eqref{A5} and also eqn.~NP3 of Ref.~\cite{ghp}, which in our basis is ($\delta_i \equiv m_i \cdot \partial$)
\be
 \delta_{[j|} \rho_{i|k]} - L_{1[j|} \rho_{i|k]} + \stackrel{l}{M}_{i[j|} \rho_{l|k]} + \stackrel{l}{M}_{[kj]} \rho_{il} = 0.
\label{A4}
\ee
The $\mu$-components of \eqref{l:n:comm} reads
\be
 D n^{\mu} = 0.
\ee
Hence
\be
\label{nmusol}
n^{\mu} = \left( n^0 \right)^{\mu},
\ee
where $n^{0\mu}$ is independent of $r$. In order to determine $n^r$, we use \eqref{A5:44} to write
\be
 L_{11} = - \frac{n^r}{r} + \rho'_{44} - \frac{\mu_0 r}{(r^2 + \chi^2)^2} + \frac{\Lambda r}{4}.
\ee
Substituting into \eqref{A5:55bar} then gives immediately
\be
\label{nreq}
 n^r = A + \left[ n^0(\chi) - F \right] \frac{r}{\chi} - \frac{\Lambda r^2}{8} - \frac{\mu_0}{2(r^2 + \chi^2)}.
\ee
The $r$-component of \eqref{l:n:comm} now gives
\be
 L_{11} = - D n^r = \frac{F - n^0(\chi)}{\chi} + \frac{\Lambda r}{4} - \frac{\mu_0 r}{(r^2 + \chi^2)^2}.
\label{L11}
\ee
All the information contained in eqn.~\eqref{A5} has now been used.

Next consider the $\mu$-components of \eqref{l:mi:comm} with $i = 4$:
\be
 D m_4^{\mu} = - \frac{m_4^{\mu}}{r}.
\ee
The solution is simply
\be
\label{m4mueq}
 m_4^{\mu} = \frac{\left( m_4^0 \right)^{\mu}}{r},
\ee
where $\left( m_4^0 \right)^{\mu}$ are functions independent of $r$. Similarly, the $r$-component gives the relation
\be
 L_{14} = - Dm_4^r - \frac{m_4^r}{r}.
\label{L14:m4r}
\ee
Substituting this into eqn.~\eqref{A4} with $ijk = 54\bar{5}$ yields
\be
 m_4^r - \frac{i m_4^0(\chi)}{r} = (r - i \chi) \left( D m_4^r + \frac{m_4^r}{r} \right) + \frac{i \chi}{r} (r - i \chi) \stackrel{4}{M}_{5\bar{5}},
\label{A4:545bar}
\ee
where $m_4^0(\chi) \equiv m_4^{0\mu} \partial_\mu \chi$. Now setting $ijk = 45\bar{5}$ in \eqref{A4}, gives
\be
 (r - i \chi) \stackrel{4}{M}_{5\bar{5}} + (r + i \chi) \stackrel{4}{M}_{\bar{5}5} = 0,
\ee
showing that $(r - i \chi) \stackrel{4}{M}_{5\bar{5}}$ is purely imaginary. The last term on the RHS of eqn.~\eqref{A4:545bar} is then real. Taking the imaginary part of \eqref{A4:545bar} gives a simple differential equation for $m_4^r$,
\be
 D m_4^r = - \frac{m_4^r}{r} + \frac{m_4^0(\chi)}{\chi r},
\ee
which integrates to
\be
\label{m4req}
 m_4^r = \frac{E_4}{r} + \frac{m_4^0(\chi)}{\chi},
\ee
for some function $E_4$ of the $x^{\mu}$-coordinates only. Going back to \eqref{L14:m4r}, one then finds
\be
 L_{14} = - \frac{m_4^0(\chi)}{\chi r}.
\label{L14}
\ee
The real part of \eqref{A4:545bar} then determines $\stackrel{4}{M}_{5\bar{5}}$,
\be
 \stackrel{4}{M}_{5\bar{5}} = - \frac{i E_4}{\chi(r - i \chi)}.
\label{M455bar}
\ee
Now following the same procedure as above determines the form of $m_5$. The $\mu$-components of \eqref{l:mi:comm} with $i = 5$ imply
\be
\label{m5musol}
 m_5^{\mu} = \frac{\left( m_5^0 \right)^{\mu}}{r + i \chi},
\ee
for complex functions $\left( m_5^0 \right)^{\mu}$ of the $x^{\mu}$-coordinates only. Then, using the $r$-component to write
\be
 L_{15} = - D m_5^r - \frac{m_5^r}{r + i \chi}
\label{L15:m5r}
\ee
and substituting into \eqref{A4} with $ijk = 55\bar{5}$ gives
\be
 D m_5^r = \frac{2 i \chi m_5^r}{r^2 + \chi^2} - \frac{i m_5^0(\chi)}{r^2 + \chi^2},
\ee
where $m_5^0(\chi) \equiv m_5^{0\mu} \partial_\mu \chi$.
Integrating this equation one finds that
\be
\label{m5rsol}
 m_5^r = \frac{E_5 (r - i \chi) + r m_5^0(\chi)}{\chi (r + i \chi)},
\ee
for some complex function $E_5$ independent of $r$. Eqn.~\eqref{L15:m5r} then determines $L_{15}$:
\be
 L_{15} = - \frac{E_5 + m_5^0 (\chi)}{\chi (r + i \chi)}.
\label{L15}
\ee
In summary, the coordinate basis components of $n^a$, $m_4^a$ and $m_5^a$ are given by (\ref{nmusol}), (\ref{nreq}), (\ref{m4mueq}), (\ref{m4req}), (\ref{m5musol}) and (\ref{m5rsol}). Thus, the $r$-dependence of the basis vectors (and hence the metric) is fully determined.

\subsection{Calculation of the non-GHP scalars $\stackrel{i}{M}_{j1}, \stackrel{i}{M}_{jk}$}

The results obtained in the previous subsection regarding the $r$-dependence of the basis vectors has automatically given information about some connection components that are not GHP scalars, namely $L_{11}$, $L_{1i}$ and $\stackrel{4}{M}_{5\bar{5}}$, as given in eqns.~\eqref{L11}, \eqref{L14}, \eqref{L15} and \eqref{M455bar}, respectively. Furthermore, our previous calculation of $\rho'_{ij}$ also provided $\stackrel{4}{M}_{51}$, eqn.~\eqref{M451}. Here we show how to obtain the other components of $\stackrel{i}{M}_{j1}, \stackrel{i}{M}_{jk}$.

We first notice that the calculation of the previous subsection does not exhaust eqn.~\eqref{A4}, there are two components remaining. The first is $ijk = 445$, for which one finds
\be
 \stackrel{4}{M}_{54} = \frac{E_5}{\chi r},
\ee
using previous results. The second is the $ijk = 545$ component, which gives
\be
 \stackrel{4}{M}_{55} = 0.
\ee

Now consider the commutator C1 of Ref.~\cite{ghp} applied to a GHP scalar $V_i$ of arbitrary boost weight $b$ and spin weight $s = 1$, $\left[ \tho, \tho' \right] V_i$. This gives two relations: a boost part (i.e. the coefficient of $b$)
\be
 D L_{11} = - \Phi + \frac{\Lambda}{4},
\ee
which is automatically satisfied, and a boost-independent part,
\be
 D \stackrel{i}{M}_{j1} = -2 \Phi^{\mathrm{A}}_{ij}.
\ee
Notice that this implies that $\stackrel{4}{M}_{51}$ is independent of $r$, which is consistent with the result already known, eqn.~\eqref{M451}. On the other hand, for $\stackrel{5}{M}_{\bar{5}1}$ this equation gives
\be
 \stackrel{5}{M}_{\bar{5}1} = i C - \frac{i \mu_0 \chi}{(r^2 + \chi^2)^2},
\label{M55bar1}
\ee
for some unknown, real function $C$ of the $x^{\mu}$-coordinates only.

One can follow the same procedure by applying the commutator C2 of \cite{ghp}, $\left[ \tho, \eth_i \right]$, to a GHP scalar $V_i$ of spin weight 1. This gives the equations
\be
 D L_{1i} = - L_{1j} \rho_{ji}
\ee
and
\be
 D \stackrel{i}{M}_{jk} = - \stackrel{i}{M}_{jl} \rho_{lk}.
\ee
The former is automatically satisfied, while the latter gives the additional information
\be
 \stackrel{5}{M}_{\bar{5}4} = \frac{i D_4}{r}
\ee
and
\be
 \stackrel{5}{M}_{\bar{5}5} = - \frac{i D_5}{r + i \chi},
\ee
where $D_4$ and $D_5$ are real and complex $r$-independent functions, respectively.

\subsection{Boost weight $-1$ components of the Weyl tensor}

We are now in a position to take $\tho'$- and $\eth_i$-derivatives in full. In particular, we can now consider the boost weight 0 components of the Bianchi equation, eqns.~B5, B5$'$, B6 and B7 of Ref.~\cite{ghp}, and determine completely $\Psi'_{ijk}$.

Start with eqn.~B5 of \cite{ghp}, which reads
\be
 - \eth_{[j|} \Phi_{i|k]} = \left( \Psi'_{[j|} \delta_{il} - \Psi'_{[j|il} \right) \rho_{l|k]}.
\label{A13}
\ee
The $ijk = 545$ component of this equation immediately gives that
\be
 \Psi'_{545} = 0.
\label{psi:p:545}
\ee
If one takes the $ijk = 55\bar{5}$ component, one finds that
\be
 \Psi'_{445} = \frac{2 \mu_0 E_5 (2 r + i \chi)(r - i \chi)}{\chi (r^2 + \chi^2)^3} + \frac{4 \mu_0 m_5^0 (\chi)}{\chi (r^2 + \chi^2)^2},
\label{psi:p:445:partial}
\ee
while setting $ijk = 54\bar{5}$ gives
\be
 \Psi'_{54\bar{5}} = - \frac{2 \mu_0 E_4 (r - i \chi)}{(r^2 + \chi^2)^3} - \frac{4 \mu_0 m_4^0(\chi)}{\chi (r^2 + \chi^2)^2}.
\ee
Substituting this result into the $ijk = 45\bar{5}$ component, one finds that
\be
 m_4^0 (\chi) = 0
\label{m40:chi}
\ee
and hence
\be
 \Psi'_{54\bar{5}} = - \frac{2 \mu_0 E_4 (r - i \chi)}{(r^2 + \chi^2)^3}.
\label{psi:p:545bar}
\ee
The only remaining independent component of \eqref{A13} is then $ijk = 445$, giving
\be
 \Psi'_{55\bar{5}} = - \frac{2 \mu_0 E_5 r (2 r - i \chi) (r - i \chi)^2}{\chi (r^2 + \chi^2)^4} - \frac{4 \mu_0 (r - i \chi)^2 m_5^0 (\chi)}{\chi (r^2 + \chi^2)^3}.
\label{psi:p:555bar:partial}
\ee
Now consider eqn.~B5$'$ of \cite{ghp},
\be
 - 2 \eth_{[j} \Phi_{k]i} + D \Psi'_{ijk} = 2 \left( \Psi'_i \delta_{[j|l} - \Psi'_{i[j|l} \right) \rho_{l|k]}.
\label{A16}
\ee
Using the previous results, the $ijk = 445$ component gives the condition
\be
m_5^0 (\chi) = - E_5.
\label{m50:chi}
\ee
Eqns.~\eqref{psi:p:445:partial} and \eqref{psi:p:555bar:partial} then imply
\be
 \Psi'_{445} = - \frac{2 i \mu_0 E_5 (r - i \chi)}{(r^2 + \chi^2)^3}
\ee
and
\be
 \Psi'_{55\bar{5}} = \frac{2 i \mu_0 E_5 (r - 2 i \chi)(r - i \chi)^2}{(r^2 + \chi^2)^4},
\ee
respectively. Of course, the symmetries of $\Psi'_{ijk}$ determine automatically
\be
 \Psi'_{45\bar{5}} = \Psi'_{54\bar{5}} - \Psi'_{\bar{5}45} = \frac{4 i \mu_0 \chi E_4}{(r^2 + \chi^2)^3}.
\ee
One can then verify that all other components of \eqref{A16} are automatically satisfied, as well as eqns.~B6 and B7 of Ref.~\cite{ghp}.

\subsection{Equation NP2$'$}

Eqn.~NP2$'$ of Ref.~\cite{ghp}, which reduces to
\be
 D \kappa'_i = \Psi'_i.
\ee
This can be integrated to determine $\kappa'_i$. The result is
\bea
 \kappa'_4 &=& G_4 + \frac{\mu_0 E_4}{(r^2 + \chi^2)^2}, \\
 \kappa'_5 &=& G_5 - \frac{i \mu_0 E_5 (r - i \chi)^2}{(r^2 + \chi^2)^3},
\eea
where $G_4$ (real) and $G_5$ (complex) are functions depending on $x^{\mu}$ only.

\subsection{Differential and algebraic constraints}

The $r$-independent integration functions appearing in the various expressions in the previous sections are not completely independent. There is still information contained in the commutators $\left[ n, m_i \right], \left[ m_i, m_j \right]$, as well as in the commutators of GHP derivatives $\left[ \tho', \eth_i \right]$ and $\left[ \eth_i, \eth_j \right]$ applied to some GHP scalar $V_i$ of spin weight 1, and this information can be used to place algebraic and differential constraints on the $r$-independent functions found above.  Later on, when we introduce coordinates, we will be interested in the symmetries and Killing fields admitted by the solutions considered here. In order to study these symmetries, we will need to know the derivatives of those functions along all basis vectors.

Consider first the commutator
\be
 \left[ m_i, m_j \right] = 2 \rho'_{[ij]} \ell + 2 \rho_{[ij]} n + 2 \stackrel{k}{M}_{[ij]} m_k
\label{mi:mj:comm}
\ee
for $i = 4, j = 5$. The $\mu$-components of this equation give
\be
 \chi \left[ m_4^0, m_5^0 \right] = - E_5 m_4^0 + \left( i \chi D_4 - i E_4 \right) m_5^0.
\label{m40:m50:comm}
\ee
On the other hand, the $r$-component can be brought to the form $p(r) = 0$, where $p(r)$ is a polynomial in $r$ with coefficients depending on the $x^{\mu}$-coordinates. It is clear that the coefficient of each power of $r$ must then vanish, resulting in the independent equations
\bea
 0 &=& \chi m_4^0 \left( E_5 \right) - 2 \chi^2 B_5 - i \chi D_4 E_5 + i E_4 E_5, \label{m4:m5:comm:a} \\
 0 &=& - i \chi m_5^0 \left( E_4 \right) + i E_4 E_5 + 3 \chi^2 B_5. \label{m4:m5:comm:b}
\eea
Since the action of $m_4^0$ and $m_5^0$ on $\chi$ is known, eqns.~\eqref{m40:chi}, \eqref{m50:chi}, one can apply their commutator \eqref{m40:m50:comm} to $\chi$ to find
\be
 m_4^0 \left( E_5 \right) = i D_4 E_5 - i \frac{E_4 E_5}{\chi}.
\ee
Comparison with \eqref{m4:m5:comm:a} yields immediately 
\be
B_5 = 0.
\ee
Hence the only non-vanishing components of $\rho'_{ij}$ are  $\rho'_{44}, \rho'_{5\bar{5}}$ (and its complex conjugate).  Using the above result, eqn.~\eqref{m4:m5:comm:b} reduces to 
\be
 m_5^0 \left( E_4 \right) = \frac{E_4 E_5}{\chi}.
\label{m50:E4}
\ee
An additional algebraic constraint can be obtained from eqn.~NP3$'$ of \cite{ghp},
\be
 \eth_{[j|} \rho'_{i|k]} = \kappa'_i \rho_{[jk]} - \frac{1}{2} \Psi'_{ijk}.
\label{A8}
\ee
By taking the $ijk = 45\bar{5}$ component and using $B_5 = 0$, one finds
\be
 G_4 = - \frac{\Lambda E_4}{4}.
\ee
Now consider $\left[ m_5, \mbar_5 \right]$ from eqn.~\eqref{mi:mj:comm}. Similarly to the calculation for $\left[ m_4, m_5 \right]$, one gets the following independent equations:
\be
 \chi \left[ m_5^0, \mbar_5^0 \right] = 2 i \chi^2 n^0 - 2 i E_4 m_4^0 + i \chi \bar{D}_5 m_5^0 + i \chi D_5 \mbar_5^0
\label{m50:m50b:comm}
\ee
and
\be
 \mbar_5^0 \left( E_5 \right) = \frac{\Lambda \chi^3}{4} +2 (A + i F) \chi - i \bar{D}_5 E_5 - \frac{E_4^2}{\chi}.
\label{m5b0:E5}
\ee
If we act with $\left[ m_5^0, \mbar_5^0 \right]$ on $\chi$ and use \eqref{m5b0:E5}, we find
\be
 n^0 (\chi) = 2F
\label{n0:chi}
\ee
and hence we know all derivatives of $\chi$.

Also, we have the commutators
\be
 \left[ n, m_i \right] = - \kappa'_i \ell + L_{1i} n - \left( \stackrel{j}{M}_{i1} + \rho'_{ji} \right) m_j.
\label{n:mi:comm}
\ee
From the $i = 4$ component one finds the relations
\be
 m_4^0 (F) = 0,
\label{m40:F}
\ee
\be
 0 = - 2 F E_4 + \chi n^0 \left( E_4 \right) - \chi m_4^0 (A),
\label{n:m4:comm:b}
\ee
and
\be
 \chi \left[ n^0, m_4^0 \right] = F m_4^0,
\label{n0:m40:comm}
\ee
while the $i = 5$ component gives
\be
 0 = m_5^0 (F) - \frac{2 F E_5}{\chi} + n^0 \left( E_5 \right) - \frac{i \Lambda E_5 \chi}{2} + G_5 \chi - i C E_5,
\label{n:m5:comm:a}
\ee
\be
 0 = \chi m_5^0 (A) - 2 i F E_5 + i \chi n^0 \left( E_5 \right) + \frac{\Lambda E_5 \chi^2}{4} - i G_5 \chi^2 + C E_5 \chi
\label{n:m5:comm:b}
\ee
and
\be
 \chi \left[ n^0, m_5^0 \right] = \left[ i C \chi + F + \frac{i \Lambda \chi^2}{4} \right] m_5^0.
\label{n0:m50:comm}
\ee
Apart from \eqref{m40:F}, which gives directly the derivative of $F$ along $m_4^0$, the other relations determine only a combination of derivatives. One can, however, gain more information from other equations.

We start by going back to eqn.~NP3$'$, \eqref{A8}. Taking the $ijk = 445$ component one finds
\be
 m_5^0 (A) = - \frac{\Lambda E_5 \chi}{4} - \frac{i F E_5}{\chi}. 
\label{m50:A}
\ee
Substituting into \eqref{n:m5:comm:b} gives
\be
 n^0 \left( E_5 \right) = G_5 \chi + i C E_5 + \frac{3 F E_5}{\chi}.
\label{n0:E5}
\ee
Now using this in \eqref{n:m5:comm:a} gives
\be
 m_5^0 (F) = \left( \frac{i \Lambda E_5}{2} - 2 G_5 \right) \chi - \frac{F E_5}{\chi}.
\label{m50:F}
\ee
Setting $ijk = 54\bar{5}$ in \eqref{A8} and using \eqref{m40:F}, one finds
\be
 m_4^0 (A) = \frac{F E_4}{\chi},
\label{m40:A}
\ee
which can now be used in eqn.~\eqref{n:m4:comm:b} to obtain
\be
 n^0 \left( E_4 \right) = \frac{3 F E_4}{\chi}.
\label{n0:E4}
\ee
No additional information can then be obtained from eqn.~NP3$'$, \eqref{A8}.

Another source of information is the commutator of GHP derivatives $\left[ \eth_i, \eth_j \right]$. Applying this to a GHP scalar $V_i$ with spin weight 1, one obtains two equations, just as in the study of $\left[ \tho, \tho' \right] V_i$ and $\left[ \tho, \eth_i \right] V_j$ carried out before. The boost part gives
\be
 \delta_{[j|} L_{1|i]} = - L_{11} \rho_{[ij]} - L_{1k} \stackrel{k}{M}_{[ij]} + \rho_{k[i|} \rho'_{k|j]} + \Phi^{\mathrm{A}}_{ij},
\ee
which is automatically satisfied using previous results. The boost-independent part is
\bea
 \delta_k \stackrel{i}{M}_{jl} - \delta_l \stackrel{i}{M}_{jk} &=& - \rho_{ik}\rho'_{jl} + \rho_{il} \rho'_{jk} + \rho_{jk} \rho'_{il} - \rho_{jl} \rho'_{ik} + 2 \stackrel{i}{M}_{j1} \rho_{[kl]} \nonumber \\
                                                               &\quad&  + \stackrel{i}{M}_{pk} \stackrel{p}{M}_{jl} - \stackrel{i}{M}_{pl} \stackrel{p}{M}_{jk} + 2 \stackrel{i}{M}_{jp} \stackrel{p}{M}_{[kl]} - \Phi_{ijkl} - \frac{\Lambda}{4} \left( \delta_{ik} \delta_{jl} - \delta_{il} \delta_{jk} \right).
\label{C3}
\eea
Setting $ijkl = 4545$ gives
\be
 m_5^0 \left( E_5 \right) = - i D_5 E_5,
\label{m50:E5}
\ee
while the $ijkl = 454\bar{5}$ gives
\be
 m_4^0 \left( E_4 \right) = - 3 F \chi.
\label{m40:E4}
\ee

There is now enough information to apply the commutators $\left[ n^0, m_i^0 \right]$, $\left[ m_i^0, m_j^0 \right]$ to the functions considered here. Before we used $\left[ m_5^0, \mbar_5^0 \right](\chi)$ to find eqn.~\eqref{n0:chi}.
Another non-trivial, algebraic relation can then be obtained by considering $\left[ n^0, m_5^0 \right](\chi)$, namely
\be
 \frac{3 i \Lambda E_5 \chi}{4} - 4 G_5 \chi - i C E_5 = 0.
\label{n0:m50:f}
\ee
Applying the same ideas to $A$, one finds that $\left[ m_4^0, m_5^0 \right] (A)$ gives
\be
 E_4 \left( G_5 - \frac{i \Lambda E_5}{4} \right) = 0.
\ee
It turns out that the term in brackets always vanishes, for suppose that $E_4 = 0$. Then \eqref{m40:E4} implies that $F = 0$, which in turn, from \eqref{m50:F}, implies
\be
 G_5 = \frac{i \Lambda E_5}{4}.
\ee
Thus, the latter is true irrespective of whether $E_4 = 0$ or $E_4 \ne 0$. Using this in \eqref{n0:m50:f}, one finds another important algebraic constraint,
\be
 E_5 \left( C + \frac{\Lambda \chi}{4} \right) = 0.
\label{C:E5}
\ee
In turn, $\left[ m_5^0, \mbar_5^0 \right] (A)$ gives
\be
 n^0 (A) = \frac{3 \Lambda F \chi}{4} + \frac{2 A F}{\chi}.
\label{n0:A}
\ee
Finally, $\left[ n^0, m_4^0 \right] (A)$ implies either $E_4 = 0$ or
\be
 n^0 (F) = \frac{2 F^2}{\chi}.
\label{n0:F}
\ee
However, as just discussed above, $E_4 = 0$ implies $F = 0$, which is consistent with \eqref{n0:F}. Therefore, one can safely take \eqref{n0:F} to be always true.

\subsection{Equation NP1$'$}

Now that we have the information about all derivatives of the functions involved, we can finally consider eqn.~NP1$'$ of Ref.~\cite{ghp} to calculate $\Omega'_{ij}$, the Weyl tensor components with boost weight $-2$. In our basis, eqn.~NP1$'$ of \cite{ghp} reads
\be
 \Delta \rho'_{ij} - \delta_j \kappa'_i = - L_{11} \rho'_{ij} + 2 \kappa'_i L_{1j} + \kappa'_k \stackrel{k}{M}_{ij} - \rho'_{kj} \stackrel{k}{M}_{i1} - \rho'_{ik} \left( \rho'_{kj} + \stackrel{k}{M}_{j1} \right) - \Omega'_{ij}.
\label{A9}
\ee
Taking the trace of this equation and recalling that $\Omega'_{ii} = 0$, one finds
\be
F = 0.
\ee
Then, the independent components of $\Omega'_{ij}$ can be obtained by putting $ij = 44, 45, 55$ in eqn.~\eqref{A9}, giving
\bea
 \Omega'_{44} &=&- \frac{4 \mu_0 \left( E_4^2 - E_5 \bar{E}_5 \right)}{(r^2 + \chi^2)^3}, \\
 \Omega'_{45} &=&  \frac{6 i \mu_0 E_4 E_5 (r - i \chi)^2}{(r^2 + \chi^2)^4}, \\
 \Omega'_{55} &=&  \frac{6 \mu_0 E_5^2 (r - i \chi)^4}{(r^2 +\chi^2)^5},
\eea
with $\Omega'_{5\bar{5}} = - \Omega'_{44}/2$ following from the traceless condition. No further information can be obtained from eqn.~\eqref{A9}.

Note that all Weyl curvature components (i.e. $\Phi_{ij}$, $\Psi'_{ijk}$ and $\Omega'_{ij}$) are proportional to $\mu_0$. Furthermore, $\Phi_{ij} \ne 0$ if $\mu_0 \ne 0$. Hence the solution is conformally flat if, and only if, $\mu_0 = 0$.

\subsection{Further algebraic constraints}

Notice that, with $F = 0$, we have
\begin{gather}
 n^0 (\chi) = 0, \qquad m_4^0 (\chi) = 0, \qquad m_5^0 (\chi) = - E_5, \notag \\[2mm]
 n^0 (A) = 0, \qquad m_4^0 (A) = 0, \qquad m_5^0 (A) = - \frac{\Lambda E_5 \chi}{4}, \notag \\[2mm]
 n^0 (E_4) = 0, \qquad m_4^0 (E_4) = 0, \qquad m_5^0 (E_4) = \frac{E_4 E_5}{\chi}.
\end{gather}
With this information, we find
\be
 \nabla_a \chi = - \frac{\bar{E}_5}{r - i \chi} \left( m_5 \right)_a - \frac{E_5}{r + i \chi} \left( \mbar_5 \right)_a.
\ee
When computing $\nabla_a A$, one then finds
\be
 \nabla_a \left( A - \frac{\Lambda \chi^2}{8} \right) = 0,
\ee
showing that the term in brackets must be constant,
\be
 A = A_0 + \frac{\Lambda \chi^2}{8}
\label{A:chi}
\ee
for some constant $A_0$. Similarly, the equation for $\nabla_a E_4$ becomes
\be
 \nabla_a (\chi E_4) = 0
\ee
and hence
\be
 E_4 = \frac{E_0}{\chi}
\label{E4:chi}
\ee
for some constant $E_0$.

We have now finished calculating all the components of the connection, including non-GHP scalars, and the components of the Weyl tensor. Furthermore, we have found the $r$-dependence of the basis vectors and several algebraic and differential constraints involving the $r$-independent functions.

\subsection{Null rotations and type D}

Consider a null rotation \eqref{null:rotation} about $\ell$. Choosing
\be
 z_4 = \frac{E_0}{\chi r}, \qquad z_5 = - \frac{i E_5}{r + i \chi},
\ee
one finds that the following GHP scalars transform according to
\be
 \kappa'_i \mapsto 0, \qquad \Psi'_{ijk} \mapsto 0, \qquad \Omega'_{ij} \mapsto 0
\ee
and
\be
 \rho'_{ij} \mapsto  b' \left(
                           \begin{array}{ccc}
                           1 + \left( a' \right)^2 & 0    & 0 \\
                           0            & 0        & 1 - i a' \\
                           0           & 1 + i a' & 0
                           \end{array}
                         \right),
\ee
where
\bea
 a' &=& \frac{\chi}{r}, \\
 b' &=& - \frac{\mu_0 r}{2(r^2 + \chi^2)^2} + \frac{A_0 r}{r^2 + \chi^2} - \frac{\Lambda r (r^2 - \chi^2)}{8 (r^2 + \chi^2)} + \frac{E_0^2}{2 r \chi^2 (r^2 + \chi^2)} + \frac{E_5 \bar{E}_5 r}{(r^2 + \chi^2)^2}.
\eea
We have thus found a basis in which only the boost weight 0 components of the Weyl tensor are non-zero, and where both $\ell$ and $n$ are geodesic with corresponding optical matrices in their canonical form. This shows that all solutions considered here are of type D.

If $b' \ne 0$ then the optical matrix of the second multiple WAND, has rank 3 (in agreement with an argument of Ref.~\cite{gspaper}). If $b' \equiv 0$ then the optical matrix of the second multiple WAND vanishes identically in which case the solution belongs to the Kundt family. The condition for this is
\be
\label{kundt}
b' \equiv 0 \qquad \Leftrightarrow \qquad  E_5 \bar{E}_5 = \frac{\mu_0}{2}, \qquad A_0 = E_0 = \Lambda=0.
\ee 
We will study this special case in more detail in section \ref{sec:newcoords}. 

\section{Coordinate basis calculations}

\label{sec:coords}

\subsection{Integrable submanifolds}

So far we have kept the coordinates $x^\mu$ arbitrary. We will now show how the results derived above lead to a canonical way of choosing these coordinates. From
\be
 \stackrel{i}{M}_{ab} = - \kappa'_i \ell_a \ell_b - \rho'_{ij} \ell_a (m_j)_b - \rho_{ij} n_a (m_j)_b + \stackrel{i}{M}_{j1} (m_j)_a \ell_b + \stackrel{i}{M}_{jk} (m_j)_a (m_k)_b,
\ee
one finds, using $B_5=0$, that
\be
 \mathrm{d} m_5 = - \left[ (\rho'_{5\bar{5}} + \stackrel{5}{M}_{\bar{5}1}) \ell + \rho_{5\bar{5}} n + (\stackrel{4}{M}_{5\bar{5}} + \stackrel{5}{M}_{\bar{5}4}) m_4 + \stackrel{5}{M}_{\bar{5}5} \bar{m_5} \right] \wedge m_5.
\ee
Thus, $m_5 \wedge \mathrm{d} m_5 = 0$. This is the integrability condition for the existence of a complex function $z$ such that $m_5 = m^{\bar{5}} \propto \mathrm{d} \bar{z}$. Thus, we write
\be
 m_5 = \bar{\cal{M}} \mathrm{d} \bar{z}
\ee
for some function $\cal{M}$. 

The distribution $\{ \ell,n,m_4 \}$ is tangent to surfaces of constant $z$ and hence integrable. This can also be seen from the commutators \eqref{l:n:comm}, \eqref{l:mi:comm}, \eqref{n:mi:comm} or, more explicitly,
\bea
 \left[ \ell, n \right]   &=& - L_{11} \ell, \\
 \left[ \ell, m_4 \right] &=& - L_{14} \ell - \rho_{44} m_4, \\
 \left[ n, m_4 \right]    &=& - \kappa'_4 \ell + L_{14} n - \rho'_{44} m_4.
\eea
We also have the commutator \eqref{n0:m40:comm} (using $F=0$)
\be
 \left[ n^0, m_4^0 \right] = 0.
\ee
Thus $\{n^0,m_4^0\}$ is integrable and we can choose coordinates $(r,u,x,z,\bar{z})$ so that
\be
 n^0 = -\frac{\partial}{\partial u} \qquad \qquad m_4^0 = N \left( \frac{\partial}{\partial x} - L_x \frac{\partial}{\partial u} \right)
\ee
for real functions $N \ne 0$,  $L_x$ independent of $r$. From $(m_5)^2=0$ we have $m_5^{\bar{z}}=0$ and hence
\be
 m_5^0 = M \left[ \frac{\partial}{\partial z} - Y \left( \frac{\partial}{\partial x} - L_x \frac{\partial}{\partial u} \right) - L_z  \frac{\partial}{\partial u} \right] 
\ee
for complex functions $M \ne 0$, $Y$, $L_z$ independent of $r$. We now have
\be
 \ell = \frac{\partial}{\partial r}, \qquad n = H \frac{\partial}{\partial r} - \frac{\partial}{\partial u},
\ee
where
\be
H \equiv  n^r = A_0 - \frac{\Lambda}{8} (r^2 - \chi^2) - \frac{\mu_0}{2(r^2 + \chi^2)}
\ee
depends on $r$. Now define $V_x$ (real) and $V_z$ (complex), both independent of $r$, by $E_4 = E_0/\chi =  -NV_x$ and $E_5 = -iM(V_z - Y V_x)$. We then have
\be
 m_4 = - \frac{NV_x}{r} \frac{\partial}{\partial r} + \frac{N}{r} \left( \frac{\partial}{\partial x} - L_x \frac{\partial}{\partial u} \right)
\ee
\be
 m_5 = - \frac{M (V_z - Y V_x)}{r + i \chi}  \frac{\partial}{\partial r} + \frac{M}{r+i\chi} \left[ \frac{\partial}{\partial z} - Y \left( \frac{\partial}{\partial x} - L_x \frac{\partial}{\partial u} \right) - L_z  \frac{\partial}{\partial u} \right].
\ee
From the inner products we then find
\bea
 \ell &=& - \d u - L_x \d x - L_z \d z - \bar{L}_z \d \bar{z}, \\
 n    &=& \d r - H \ell + V_x \d x + V_z \d z + \bar{V}_z \d \bar{z}, \\
 m_4  &=& \frac{r}{N} \left( \d x + Y \d z + \bar{Y} \d \bar{z} \right), \\
 m_5  &=& \frac{r - i \chi}{\bar{M}} \d \bar{z}
\eea
from which it is easy to write down the metric.

Recall that $n^0(\chi) = m_4^0(\chi) = 0$. These imply that
\be
 \partial_u \chi = \partial_x \chi = 0,
\ee
hence $\chi = \chi(z,\bar{z})$. Now, $m_5^0(\chi) = - E_5$ reduces to
\be \label{eqn:E5}
 \partial_z \chi = - \frac{E_5}{M},
\ee
while the commutator $\left[ n^0, m_4^0 \right] = 0$ gives
\be
 \partial_u N = \partial_u L_x = 0.
\ee
From \eqref{E4:chi} we then have $\partial_u V_x = 0$. The result above implies that $L_x = L_x(x,z,\bar{z})$. Note that we have the residual coordinate freedom $u \rightarrow u' = u + h(x,z,\bar{z})$, which has the effect
\be
 L_x \rightarrow L_x' = L_x - \partial_x h, \qquad L_z \rightarrow L_z' = L_z - \partial_z h.
\ee
We can therefore choose $h$ appropriately to set $L_x' = 0$, and then drop the primes. Henceforth,
\be
 L_x = 0.
\ee
The expression
\be
 \left[ n^0, m_5^0 \right] = i \left( C + \frac{\Lambda \chi}{4} \right) m_5^0
\ee
reduces to
\be
 \partial_u L_z = \partial_u Y = 0, \qquad \partial_u M = - i \left( C + \frac{\Lambda \chi}{4} \right) M.
\ee
The latter equation implies
\be
 \partial_u |M|^2 = 0,
\ee
and
\be
 C = - \frac{\Lambda \chi}{4} + \frac{i}{2} \left[ \frac{n^0(\bar{M})}{\bar{M}} - \frac{n^0(M)}{M} \right],
\ee
where we used the fact that $\partial_u M/M$ is purely imaginary to write $C$ in a form that is manifestly real. If we now recall that $n^0(E_5) = 0$, we find
\be
 0 = \partial_u E_5 = - i M \partial_u V_z,
\ee
where we make use of \eqref{C:E5}. Hence, $\partial_u V_z = 0$. Thus, we conclude that all functions appearing in the metric
\be
 g_{ab} = \ell_a n_b + n_a \ell_b + (m_i)_a (m_i)_b
\label{metric:covectors}
\ee
are independent of $u$, and hence $\partial/\partial u$ is a Killing vector field. Notice, in particular, that the metric depends on $M$ only through $|M|^2$.

The commutator
\be
 \left[ m_4^0, m_5^0 \right] = - \frac{E_5}{\chi} m_4^0 + i \left( D_4 - \frac{E_0}{\chi^2} \right) m_5^0
\ee
gives
\bea
 \partial_x L_z &=& \partial_z L_x = 0, \\
 \partial_z (\chi N) &=& \chi Y \partial_x N - \chi N \partial_x Y, \label{coords:m40:m50:b} \\
 m_4^0(M) &=& i M \left( D_4 - \frac{E_4}{\chi} \right).
\eea
The latter equation can be used to show that
\be
 \partial_x |M|^2 = 0,
\ee
as well as to write $D_4$ as
\be
 D_4 = - \frac{N V_x}{\chi} + \frac{i}{2} \left[ \frac{m_4^0(\bar{M})}{\bar{M}} - \frac{m_4^0(M)}{M} \right].
\ee
Moreover, there is still freedom in redefining $x$ by transforming $x \rightarrow x'(x,z,\bar{z})$. This changes $N$ to $N' = N \partial x'/\partial x$. Since $N = N(x,z,\bar{z})$, we can use this to impose the condition $N' = 1/\chi$. Dropping the primes, $N = 1/\chi$, which implies that the LHS of \eqref{coords:m40:m50:b} vanishes, as well as $\partial_x N$. Thus \eqref{coords:m40:m50:b} simply reduces to
\be
\partial_x Y = 0.
\ee
Notice also that $E_4 = E_0/\chi = - N V_x$ implies that $V_x = -E_0 = \mathrm{constant}$. The equation for $m_4^0(E_5)$,
\be
 m_4^0 (E_5) = i D_4 E_5 - \frac{i E_0 E_5}{\chi^2},
\ee
then reduces to
\be
 \partial_x V_z = 0.
\ee
Therefore, similarly to the argument above for $\partial/\partial u$, we have shown that $\partial/\partial x$ is also a Killing vector field for these solutions. The function $M$ depends on $u$ and $x$ only through a phase, which may be eliminated with an $r$-independent spin transformation of the form \eqref{spin}.

Consider now the expression for $\left[ m_5^0, \mbar_5^0 \right]$,
\be
 \left[ m_5^0, \mbar_5^0 \right] = 2 i \chi n^0 - \frac{2 i E_0}{\chi^2} m_4^0 + i \bar{D}_5 m_5^0 + i D_5 \mbar_5^0,
\ee
which reduces to
\bea
 2i\chi &=& |M|^2 \left( \partial_z \bar{L}_z - \partial_{\bar{z}} L_z \right), \label{Lz:eqn} \\
 - \frac{2i E_0}{\chi^3} &=& |M|^2 \left( \partial_{\bar{z}} Y - \partial_z \bar{Y} \right), \label{Y:eqn} \\
 D_5 &=& - \frac{i m_5^0(\bar{M})}{\bar{M}}.
\eea
With the above equation for $D_5$, the equation
\be
 m_5^0(E_5) = -i D_5 E_5
\ee
can be written as
\be
 m_5^0(E_5 \bar{M}) = 0.
\ee
But if we recall that $E_5 = -M \partial_z \chi$, see \eqref{eqn:E5}, the quantity in brackets is $-|M|^2 \partial_z \chi$, which is a function of $z,\bar{z}$ only. The equation above then shows that it must actually be a function of $\bar{z}$ only:
\be
 E_5 \bar{M} = - |M|^2 \partial_z \chi = \bar{k},
\ee
for some analytic function $k = k(z)$. If we now consider a holomorphic transformation $z \rightarrow z'(z)$, the effect is to change $M \rightarrow M' = M d z'/d z$ and hence $k \rightarrow k' = k d z'/d z$. We can therefore use this transformation to set $k'$ to $-1$ or $0$. Dropping the primes, we have shown that we can write
\be
 E_5 \bar{M} = - |M|^2 \partial_z \chi = k_0 = \mathrm{constant} \in \{-1,0\}.
\label{E5:M:k0}
\ee
We now see that $\partial_z \chi = \partial_{\bar{z}} \chi$. If we now write $z = z_1 + i z_2$, we see that
\be
 \partial_2 \chi \equiv \frac{\partial \chi}{\partial z_2} = 0,
\ee
hence $\chi = \chi(z_1)$. We can now distinguish between two different cases.

\subsection{Case 1: $d\chi \ne 0$}

We will now prove case 1 of our Theorem. $d \chi \ne 0$ implies $\partial_z \chi  \ne 0$ so we must have $k_0 = -1$. The expression
\be
 \mbar_5^0 (E_5) = 2 A_0 \chi + \frac{\Lambda \chi^3}{2} - i \bar{D}_5 E_5 - \frac{E_0^2}{\chi^3},
\label{m5b:E5}
\ee
which is derived from eqn.~\eqref{m5b0:E5} and previous results, reduces to
\be
 - |M|^2 \frac{\partial^2 \chi}{\partial z \partial \bar{z}} = 2 A_0 \chi + \frac{\Lambda \chi^3}{2} - \frac{E_0^2}{\chi^3}.
\ee
Using $\partial_z \chi = \partial_1 \chi/2 = 1/|M|^2$, we can rewrite this as
\be
 \partial_1 \left( \partial_1 \chi + \frac{E_0^2}{\chi^2} + 2 A_0 \chi^2 + \frac{\Lambda \chi^4}{4} \right) = 0.
\ee
Since the quantity in brackets is a function of $z_1$ only, it follows that it must be a constant $C_0$. Thus,
\be
 \frac{d\chi}{dz_1} =   C_0 - \frac{E_0^2}{\chi^2} - 2 A_0 \chi^2 - \frac{\Lambda \chi^4}{4} \equiv P(\chi).
\label{P:def}
\ee
We can therefore use $\chi$ as a coordinate rather than $z_1$, with the transformation rule given by \eqref{P:def}. Note that $\partial_1 \chi = 2/|M|^2$ implies
\be
 |M|^2 = \frac{2}{P(\chi)}
\ee
from which it follows that $\chi$ must lie in a range for which $P(\chi)>0$. 

The only quantities that we do not yet know are the functions $L_z$ and $Y$, which are determined by eqns.~\eqref{Lz:eqn} and \eqref{Y:eqn}, respectively. Both have the same structure:
\be
 |M|^2 \left( \partial_z \bar{\mathcal{F}} - \partial_{\bar{z}} \mathcal{F} \right) = \mathcal{G}(\chi),
\ee
where ${\cal F}$ denotes $L_z$ or $Y$ and the RHS is a known function. Notice that
\be
 \d (\mathcal{F} \d z + \bar{\mathcal{F}} \d \bar{z}) = \left( \partial_z \bar{\mathcal{F}} - \partial_{\bar{z}} \mathcal{F} \right) \d z \wedge \d \bar{z}
\ee
and the quantity in brackets on the LHS is precisely the combination in which both $L_z$ and $Y$ appear in the metric. If a particular solution $\mathcal{F} \d z + \mathcal{F} \d \bar{z}$ is found, any other solution will differ from this by a gradient $\d \alpha$, where $\alpha = \alpha(z,\bar{z})$. In the case of $L_z$, this can be absorbed by defining a new coordinate $u' = u - \alpha(z,\bar{z})$, which does not change any other quantity and, in particular, $L_x = 0$ is maintained. Similarly, defining $x' = x - \alpha(z,\bar{z})$ eliminates the gradient $\d \alpha$ from the expression for $Y$ (notice that $\partial x'/\partial x = 1$, and hence every other quantity is unchanged). Hence all we need to do is to find particular solutions for $L_z$ and $Y$. 

Consider first the equation for $L_z$. If we search for a particular solution in which $L_z$ depends only on $\chi$, \eqref{Lz:eqn} becomes
\be
 \frac{d \bar{L}_z}{d \chi} - \frac{d L_z}{d \chi} = 2 i \chi.
\ee
This can be solved, in particular, for
\be
 L_z = - \frac{i \chi^2}{2}.
\ee
Similarly, if we look for a solution for $Y$ such that $Y = Y(\chi)$, \eqref{Y:eqn} reduces to
\be
 \frac{d Y}{d \chi} - \frac{d \bar{Y}}{d \chi} = -\frac{2 i E_0}{\chi^3}.
\ee
A solution is
\be
 Y = \frac{i E_0}{2 \chi^2}.
\ee
Then, from the definition $E_5 = - i M(V_z - Y V_x)$ and from \eqref{E5:M:k0}, \eqref{P:def}, one finds
\be
 V_z = - \frac{i}{2} \left( P + \frac{E_0^2}{\chi^2} \right) = - \frac{i}{2} \left( C_0 - 2 A_0 \chi^2 - \frac{\Lambda \chi^4}{4} \right).
\ee
Thus, all functions appearing in the metric \eqref{metric:covectors} depend only on $\chi$ (equivalently $z_1$) and $r$. In addition to $\partial/\partial u$ and $\partial/\partial x$, we now find that $\partial/\partial z_2$ is also a Killing vector field for this metric. Using our definitions and results above, the metric becomes
\bea
 ds^2 &=& - 2 (\d u + \chi^2 \d z_2) \left[ \d r + H (\d u + \chi^2 \d z_2) - E_0 \left( \d x - \frac{E_0}{\chi^2} \d z_2 \right) + P \d z_2 \right] \nonumber \\
      &\qquad& + r^2 \chi^2 \left( \d x - \frac{E_0}{\chi^2} \d z_2 \right)^2 + (r^2 + \chi^2)  \left( \frac{\d \chi^2}{P} + P \d z_2^2 \right).
\eea
Making the definition $y \equiv z_2$ this gives the metric (\ref{unequal}) in the statement of our Theorem.

\subsection{Case 2: $d\chi \equiv 0$}

In this case, $\chi$ is constant. Eqn.~(\ref{E5:M:k0}) gives $E_5 =k_0 = 0$. The equation for $m_5^0(E_5)$ is then trivial, but the one for $\mbar_5^0 (E_5)$, eqn.~\eqref{m5b:E5}, becomes
\be
 A_0 = \frac{E_0^2}{2\chi^4} - \frac{\Lambda}{4} \chi^2,
\label{chiconst-constraint}
\ee
which gives an algebraic constraint involving these constants. There is a non-trivial component of \eqref{C3} that has not yet been considered. The $ijkl = 5\bar{5}5\bar{5}$ component gives
\be
 0 = i \chi^2 \left[ m_5^0 \left( \bar{D}_5 \right) - \mbar_5^0 \left( D_5 \right) \right] + 2 A_0 \chi^2 + \Lambda \chi^4 - 2 C \chi^3 + \frac{E_0^2}{\chi^2} + 2 E_0 D_4 + 2 D_5 \bar{D_5} \chi^2.
\ee
In general, using our expressions for $C, D_4, D_5$, this gives a second-order differential equation involving $M, \bar{M}$ which can be put in the form (using (\ref{chiconst-constraint}))
\be
 R^{(2)} = \frac{8 E_0^2}{\chi^4} + 2 \Lambda \chi^2,
\ee
where $R^{(2)}$ is the Ricci scalar of the two-dimensional metric
\be
 g^{(2)} = \frac{2}{|M|^2} \d z \d \bar{z}.
\label{2dim-metric}
\ee
Hence this two-dimensional metric has constant curvature. A volume form for this two-dimensional metric is
\be
 \epsilon^{(2)} = \frac{i}{|M|^2} \d z \wedge \d \bar{z}.
\ee
Eqns.~\eqref{Lz:eqn} and \eqref{Y:eqn} then become
\bea
 \d (L_z \d z + \bar{L}_z \d \bar{z}) &=& 2 \chi \epsilon^{(2)}, \\
 \d (Y \d z + \bar{Y} \d \bar{z}) &=& \frac{2 E_0}{\chi^3} \epsilon^{(2)},
\eea
respectively. If we define a one-form $\mathcal{A}={\cal A}_z(z,\bar{z}) dz + {\cal A}_{\bar{z}}(z,\bar{z}) d\bar{z}$ by
\be
 \epsilon^{(2)} = \d \mathcal{A},
\label{Acal:def}
\ee
then particular solutions to the equations above are
\be
 L_z \d z + \bar{L}_z \d \bar{z} = 2 \chi \mathcal{A}, \qquad Y \d z + \bar{Y} \d \bar{z} = \frac{2 E_0}{\chi^3} \mathcal{A}.
\ee
As in case 1, any other solution would differ from these by some gradients $\d \alpha(z,\bar{z})$ and $\d \beta(z,\bar{z})$, respectively. These can be absorbed into $u$ and $x$ using the residual coordinate freedom $u' = u + \alpha(z,\bar{z})$, $x' = x + \beta(z,\bar{z})$, which preserve all quantities fixed above. Then, from $E_5 = - i M(V_z - Y V_x) = 0$ one finds $V_z = - E_0 Y$, and hence
\be
 V_z \d z + \bar{V}_z \d \bar{z} = - \frac{2 E_0^2}{\chi^3} \mathcal{A}.
\ee
The metric is then
\bea
\label{constantchi}
 ds^2 &=& - 2 (\d u + 2 \chi \mathcal{A}) \left[ \d r + H (\d u + 2 \chi \mathcal{A}) - E_0 \left( \d x + \frac{2 E_0}{\chi^3} \mathcal{A} \right) \right] \nonumber \\
      &\qquad& + r^2 \chi^2 \left( \d x + \frac{2 E_0}{\chi^3} \mathcal{A} \right)^2 + (r^2 + \chi^2) g^{(2)}.
\eea
We now transform to arbitrary real coordinates $y^\alpha(z,\bar{z})$ on the 2d part of the metric, so that $g^{(2)} = h_{\alpha \beta}(y) dy^\alpha dy^\beta$, and this gives the metric (\ref{equal}) in the statement of our Theorem.

The metric (\ref{constantchi}) has a scaling symmetry analogous to (\ref{scaling}). For $\lambda \ne 0$ we can perform the coordinate transformation
\be
u = \frac{u'}{\lambda} \qquad r = \lambda r' \qquad  x = \frac{x'}{\lambda^2}
\ee
and the metric takes the same form as before but now with the constants rescaled as 
\be
\label{scaling2}
\chi' = \frac{\chi}{\lambda}  \qquad \mu_0' = \frac{\mu_0}{\lambda^4}  \qquad E_0' = \frac{E_0}{\lambda^3}
\ee
and $g^{(2)}$ replaced by $g^{(2)'} = \lambda^2 g^{(2)}$ and $\mathcal{A}$ replaced by $\mathcal{A}' = \lambda^2 \mathcal{A}$. 

\section{Relation to Kerr-de Sitter}

In this section we will perform coordinate transformations to demonstrate how the metrics (\ref{unequal}) and (\ref{equal}) are related to the 5d Kerr-de Sitter solution. 

\label{sec:newcoords}

\subsection{Case 1: $d\chi \ne 0$}

\label{unequalcoords}

If we define the 1-forms
\be
 \sigma^1 = \d u + \chi^2 \d y, \qquad \sigma^2 = \d x - \frac{E_0}{\chi^2} \d y, \qquad \sigma^3 = \d y
\ee
then the metric (\ref{unequal}) can be written
\be
 ds^2 = -2 \sigma^1 \d r + \frac{r^2+\chi^2}{P} \d \chi^2 + h_{ij} \sigma^i \sigma^j,
\ee
where
\be
 h_{ij} = \left( \begin{array}{lll}  -2H & E_0 & -P \\ E_0 & r^2 \chi^2 & 0\\  -P & 0 & (r^2+\chi^2) P \end{array} \right).
\ee
If we let $x^I = \{u,x,y\}$ then the metric can be written
\be
 ds^2 = -2a_I \d x^I \d r + g_{IJ} \d x^I \d x^J + \frac{r^2+\chi^2}{P} \d \chi^2,
\ee
where
\be
 a_I = \sigma^1_I, \qquad g_{IJ} = h_{ij} \sigma^i_I \sigma^j_J.
\ee
Now consider a change of coordinates 
\be
 \d x^I = \d y^I + A^I(r) \d r
\ee
for some functions $A^I(r)$. We want to choose the functions $A^I$ to eliminate $\d y^I \d r$ terms from the resulting metric. This requires
\be
\label{Aa}
 A^I = g^{IJ} a_J.
\ee
We can only do this if the RHS above is independent of $\chi$. We find that\footnote{To do this computation it is convenient to define $\eta_1 = \partial/\partial u$, $\eta_2 = \partial/\partial x$ and $\eta_3 = \partial/\partial z_2 - \chi^2 \partial/\partial u + (E_0/\chi^2) \partial/\partial x$ so that $\sigma^i (\eta_j) = \delta^i_j$. We then have $g^{IJ} = h^{ij} \eta_i^I \eta_j^J$ where $h^{ij}$ is the inverse of $h_{ij}$.}
\be
 A^1 = -\frac{r^2}{F(r)}, \qquad A^2 = \frac{E_0}{ r^2 F(r)}, \qquad A^{3} = -\frac{1}{F(r)},
\ee
where
\bea
\label{Fdef}
 F( r) &=& \left( r^2 + \chi^2 \right) \left( 2H + \frac{E_0^2}{r^2 \chi^2} + \frac{P}{r^2 + \chi^2} \right) \nonumber \\
 &=& -\frac{\Lambda}{4} r^4 + 2A_0 r^2 - \mu_0 + C_0 + \frac{E_0^2}{r^2}.
\eea
So the RHS of (\ref{Aa}) is indeed independent of $\chi$ and the coordinate transformation is permissible provided we work in a region where $F(r ) \ne 0$. Note that
\be
 F( r) = P(ir ) - \mu_0.
\ee
The metric in the new coordinates is
\be
  ds^2 = \left( r^2 + \chi^2 \right) \left[ \frac{\d r^2}{F( r)} + \frac{\d \chi^2}{P(\chi)} \right] + h_{ij} \nu^i \nu^j,
\ee
where $\nu^i$ is defined by replacing $x^I$ with $y^I$ in  $\sigma^i$. Now using (\ref{Fdef}) to eliminate $H$ we can write 
\be
 h_{ij} \nu^i \nu^j = -\frac{F( r)}{r^2 + \chi^2} (\nu^1)^2 + \frac{P(\chi)}{r^2 + \chi^2} \left[ \nu^1 - (r^2 + \chi^2) \nu^3 \right]^2 + \frac{1}{r^2 \chi^2} \left[ E_0 \nu^1 + r^2 \chi^2 \nu^2 \right]^2 .
\ee
For $E_0 \ne 0$ we define 
\be
 y^1 = \psi_0, \qquad y^2 = E_0 \psi_2, \qquad y^3 = \psi_1,
\ee
and the metric is
\bea
  ds^2 &=& \left( r^2 + \chi^2 \right) \left[ \frac{\d r^2}{F( r)}+ \frac{\d \chi^2}{P(\chi)} \right]   -\frac{F( r)}{r^2 + \chi^2} (\d \psi_0 + \chi^2 \d \psi_1 )^2\nonumber \\ &+& \frac{P(\chi)}{r^2 + \chi^2} \left[ \d \psi_0 - r^2 \d \psi_1 \right]^2 + \frac{E_0^2}{r^2 \chi^2} \left[ \d \psi_0 + (\chi^2 - r^2) \d \psi_1 + r^2 \chi^2 \d \psi_2 \right]^2 .
\eea
If we now define $x_1 = \chi$ and $x_2 = i r$ then this is the Kerr-de Sitter solution \cite{hht} with two non-zero, unequal, spin parameters, as written in eqn.~(22) of Ref.~\cite{chen}.\footnote{
The parameters of Ref.~\cite{chen} are given in terms of our parameters by $c_1 = 2A_0$, $c_2 = \Lambda/4$, $c = E_0^2$, $b_1 = C_0/2$, $b_2 = C_0/2 - \mu_0/2$.}

The case $E_0=0$ corresponds to the Kerr-de Sitter metric with a single non-vanishing spin parameter. It can be obtained from the solution as written in Ref.~\cite{chen} by defining $\psi_2 = \hat{\psi_2}/E_0$ and taking the limit $E_0 \rightarrow 0$. 

Now we return to the special case $F( r) \equiv 0$, for which the above coordinate transformation no longer works. This condition can be understood geometrically as follows. The metric admits a 3d isometry group (associated to the Killing fields $\partial/\partial u$, $\partial/\partial x$, $\partial/\partial y$) with 3d orbits. The condition $F( r) \equiv 0$ is the condition for these orbits to be null everywhere.\footnote{This is analogous to the 4d solution given in eqn.~(24.21) of \cite{exactsolutions}, which can be obtained as a limit of the Kerr-NUT solution.}  Note that
\be
 F( r) \equiv 0 \qquad \Leftrightarrow \qquad \Lambda = A_0 = E_0 = 0 \qquad C_0 = \mu_0.
\ee
This is precisely the condition (\ref{kundt}) for the second multiple WAND to have vanishing expansion, rotation and shear, i.e., for the spacetime to be Kundt.\footnote{To see this, note that (\ref{E5:M:k0}) implies $|E_5|^2 = 1/|M|^2$ and we have $1/|M|^2 = P(\chi)/2 = C_0/2$ using $A_0 = E_0 = 0$.}

In this special case, the metric simplifies to
\be
\label{specialMP}
 ds^2 = - 2dr \left( du + \chi^2 dy \right) + \frac{\mu_0}{r^2+\chi^2} \left( du - r^2 dy \right)^2 + r^2 \chi^2 dx^2 + \frac{r^2+\chi^2}{\mu_0} d\chi^2,
\ee
where $\mu_0>0$ follows from $P(\chi)>0$. (The scaling symmetry (\ref{scaling}) could be used to set $\mu_0=1$.) For this metric, the second multiple WAND is
\be
 k = \frac{1}{r^2 + \chi^2} \left[ r^2 \frac{\partial}{\partial u} + \frac{\partial}{\partial y} \right],
\ee
which obeys $k_a = -(dr)_a$ so surfaces of constant $r$ are all null. A surface $r=r_0$ is a Killing horizon of the vector field $r_0^2 \partial/\partial u +  \partial/\partial y$. The surface gravity vanishes, so this spacetime is foliated by degenerate Killing horizons. The metric is smooth at $\chi=0$ (for $r \ne 0$) provided that $x$ is identified with period $2\pi/\sqrt{\mu_0}$, which implies that cross-sections of the Killing horizons have topology $\mathbb{R}^3$ (assuming the coordinates $u,y$ are non-compact). The curvature diverges at $r=\chi=0$ (see \eqref{Phi}). The curvature vanishes as $r \rightarrow \infty$ so the spacetime is asymptotically locally flat.

Obviously this special case is a limit of the generic case with $F(r ) \ne 0$. One can obtain the solution as a limit of a regular Myers-Perry black hole solution. Start from a single spinning Myers-Perry black hole, which has $\Lambda=E_0=0$ and $A_0>0$ and $C_0>0$. Using the scaling freedom (\ref{scaling}) we set $A_0 =1/2$. Then the MP spin parameter is $a=\sqrt{C_0}$ and the mass parameter is $\mu_0$. It has a regular horizon provided $\mu_0>a^2$. The extremal solution with $\mu_0=a^2$ is nakedly singular. To take the limit, perform the rescaling (\ref{scaling}) and take the limit $\lambda \rightarrow \infty$ with $\mu_0/\lambda^4$ fixed and $(\mu_0-a^2)/\lambda^4 \rightarrow 0$. This corresponds to scaling the Myers-Perry black hole towards the extremal solution whilst simultaneously taking its mass to infinity.\footnote{The limit involves ``zooming in" on the equatorial plane of the black hole, inside the ergoregion, which explains why the limiting metric is non-stationary.}

\subsection{Case 2: $d\chi \equiv 0$}

In the metric (\ref{equal}) written as in (\ref{constantchi}), set
\be
 x= \frac{x'}{\chi} + \frac{E_0}{\chi^4} u
\ee
and let
\be
 \sigma = \d u + 2 \chi {\cal A}.
\ee
The metric becomes
\bea
 ds^2 = - 2\sigma  \d r - G(r) {\sigma}^2 + r^2  \left[ \d x' + \frac{E_0}{\chi^3} \left( 1 + \frac{\chi^2}{r^2} \right) \sigma \right]^2 + (r^2 + \chi^2) g^{(2)},
\eea
where
\bea
G(r) &=& 2H(r) + \frac{E_0^2}{r^2 \chi^2}\nonumber \\  &=& \frac{E_0^2}{\chi^4} \left( 1 + \frac{\chi^2}{r^2} \right) - \frac{\Lambda}{4} (r^2 + \chi^2) - \frac{\mu_0}{r^2 + \chi^2}.
\eea 
Now let
\be
 \d u = \d p + A(r) \d r, \qquad \d x' = \d q + B(r) \d r
\ee
and choose $A(r)$, $B(r)$ to eliminate $\d p \d r$ and $\d q  \d r$ terms from the metric. This gives\footnote{
Here we assume that we are working in a region with $G(r) \ne 0$. If $G(r) \equiv 0$ then $\mu_0 = 0$ in which case the Weyl tensor vanishes so the solution is conformally flat.}
\be
A(r) = -\frac{1}{G(r)}, \qquad B(r) = \left( 1 + \frac{\chi^2}{r^2} \right) \frac{E_0}{\chi^3 G(r)}
\ee
and the metric becomes
\be
\label{constantchi2}
 ds^2 = \frac{\d r^2}{G(r)} - G(r) \nu^2 + r^2 \left[\d q + \frac{E_0}{\chi^3} \left( 1 + \frac{\chi^2}{r^2} \right) \nu \right]^2+ (r^2 + \chi^2) g^{(2)},
\ee
where
\be
 \nu = \d p + 2\chi {\cal A}.
\ee
The local symmetries of $g^{(2)}$, which has constant curvature, extend to symmetries of the full metric (\ref{constantchi2}), hence this metric is cohomogeneity-1, where the surfaces of (local) homogeneity are surfaces of constant $r$. 

If $E_0=0$ then the Killing vector field $\partial/\partial q$ is hypersurface orthogonal. The scaling symmetry (\ref{scaling2}) can be used to eliminate one parameter, so these form a 1-parameter family. These solutions were discussed in Ref.~\cite{mann03}. 

To analyse the metric (\ref{constantchi2}), consider the case for which $R^{(2)} > 0$. Perform the coordinate transformation
\be
\label{ppsi}
 p = \chi \psi   - \frac{8 E_0}{\chi R^{(2)}} q,
\ee
where the coefficient of the second term is chosen to eliminate the ${\cal O}(r^2)$ terms in the $g_{q\psi}$ component of the resulting metric. After defining new coordinates $\rho = \sqrt{r^2 + \chi^2}$ and $t = \chi q$, the metric is
\be
\label{cohom1mp}
 ds^2 = -\frac{\beta(\rho)}{\gamma(\rho)} \d t^2 + \frac{\d \rho^2}{\beta(\rho)} +\rho^2 \gamma(\rho) \left( \d \psi + 2 {\cal A} - \Omega(\rho) \d t \right)^2 + \rho^2 g^{(2)},
\ee
where
\be
 \beta(\rho) = \left(-\frac{\Lambda}{4} \rho^2 + \frac{R^{(2)}}{8} - \frac{\mu_0}{\rho^2} + \frac{\mu_0 \chi^2}{\rho^4} \right),
\ee
\be
 \gamma(\rho) = \frac{R^{(2)}}{8}  + \frac{\mu_0 \chi^2}{\rho^4}
\ee
and
\be
 \Omega(\rho) = \frac{8 \mu_0 E_0}{\rho^4 \gamma(\rho) R^{(2)}}.
\ee
We can use the scaling freedom (\ref{scaling2}) to set $R^{(2)}=8$ so (\ref{const:curv}) implies $\Lambda \chi^2/4 \le 1$. If this inequality is strict (which is always the case for $\Lambda \le 0$) then the above metric is the Kerr-de Sitter solution with equal rotation parameters as written in \cite{klr}, where the 2 parameters (spin and mass) are
\be
 a = \frac{\chi}{\sqrt{1- \frac{\Lambda}{4} \chi^2}}, \qquad M = \frac{\mu_0}{2}\left( 1 -\frac{\Lambda}{4} \chi^2 \right).
\ee
If the inequality is saturated, i.e., $\Lambda \chi^2/4=1$ (only possible for $\Lambda>0$) then we have $E_0=0$. We see that this solution can be obtained by taking a limit of the Kerr-de Sitter solution in which $a \rightarrow \infty$, $M \rightarrow 0$ with $M a^2$ approaching a finite limit.

If $\Lambda>0$ then $R^{(2)}>0$ is the only possibility. If $\Lambda=0$ then it is also possible to have $R^{(2)}=0$, which requires $E_0=0$. If $\Lambda<0$ then we can have $R^{(2)}=0$ (with $E_0 \ne 0$) or $R^{(2)}<0$. Note that the coordinate transformation (\ref{ppsi}) that brings the metric to the form (\ref{cohom1mp}) is valid also if $R^{(2)}<0$. 

\subsection*{Acknowledgment}

This work was supported by the European Research Council grant no.\ ERC-2011-StG 279363-HiDGR. G.B.F.\ is supported by CAPES grant no. 0252/11-5.  M.G.\ is supported by King's College, Cambridge. We are grateful to Roberto Emparan for a discussion of the solution (\ref{specialMP}).


\begin{thebibliography}{99}

\bibitem{exactsolutions}
H.~Stephani, D.~Kramer, M.~A.~H.~MacCallum, C.~Hoenselaers and E.~Herlt,
  ``Exact solutions of Einstein's field equations,'' 2nd edition, 
  Cambridge Univ. Pr. (2003) 

\bibitem{kinnersley} 
  W.~Kinnersley,
  J.\ Math.\ Phys.\  {\bf 10}, 1195 (1969).

\bibitem{cmpp}
A.~Coley, R.~Milson, V.~Pravda and A.~Pravdova,
  Class.\ Quant.\ Grav.\  {\bf 21}, L35 (2004)
  [gr-qc/0401008].

\bibitem{ortaggio}
M.~Ortaggio,
  Class.\ Quant.\ Grav.\  {\bf 26}, 195015 (2009)
  [arXiv:0906.3818 [gr-qc]].

\bibitem{myersperry}
R.~C.~Myers and M.~J.~Perry,
  Annals Phys.\  {\bf 172}, 304 (1986).

\bibitem{frolov}
V.~P.~Frolov and D.~Stojkovic,
  Phys.\ Rev.\ D {\bf 68}, 064011 (2003)
  [gr-qc/0301016].

\bibitem{hourityped}
N.~Hamamoto, T.~Houri, T.~Oota and Y.~Yasui,
  J.\ Phys.\ A {\bf 40}, F177 (2007)
  [hep-th/0611285].

\bibitem{typed}
V.~Pravda, A.~Pravdova and M.~Ortaggio,
  Class.\ Quant.\ Grav.\  {\bf 24}, 4407 (2007)
  [arXiv:0704.0435 [gr-qc]].

\bibitem{hht}
S.~W.~Hawking, C.~J.~Hunter and M.~Taylor,
  Phys.\ Rev.\ D {\bf 59}, 064005 (1999)
  [hep-th/9811056].


\bibitem{glpp}
G.~W.~Gibbons, H.~Lu, D.~N.~Page and C.~N.~Pope,
  J.\ Geom.\ Phys.\  {\bf 53}, 49 (2005)
  [hep-th/0404008].

\bibitem{geodesic}
M.~Durkee and H.~S.~Reall,
  Class.\ Quant.\ Grav.\  {\bf 26}, 245005 (2009)
  [arXiv:0908.2771 [gr-qc]].


\bibitem{chen}
W.~Chen, H.~Lu and C.~N.~Pope,
  Class.\ Quant.\ Grav.\  {\bf 23}, 5323 (2006)
  [hep-th/0604125].


\bibitem{bubble}
F.~Dowker, J.~P.~Gauntlett, G.~W.~Gibbons and G.~T.~Horowitz,
  Phys.\ Rev.\ D {\bf 52}, 6929 (1995)
  [hep-th/9507143].

\bibitem{gspaper}
 M.~Ortaggio, V.~Pravda, A.~Pravdova and H.~S.~Reall,
  Class.\ Quant.\ Grav.\  {\bf 29}, 205002 (2012)
  [arXiv:1205.1119 [gr-qc]].

\bibitem{morisawa}
Y.~Morisawa and D.~Ida,
  Phys.\ Rev.\ D {\bf 69}, 124005 (2004)
  [gr-qc/0401100].
  
\bibitem{hollands}
S.~Hollands and S.~Yazadjiev,
  Commun.\ Math.\ Phys.\  {\bf 283}, 749 (2008)
  [arXiv:0707.2775 [gr-qc]].

\bibitem{houri}
T.~Houri, T.~Oota and Y.~Yasui,
  Phys.\ Lett.\ B {\bf 656}, 214 (2007)
  [arXiv:0708.1368 [hep-th]].
  
 \bibitem{krtous}
  P.~Krtous, V.~P.~Frolov and D.~Kubiznak,
  Phys.\ Rev.\ D {\bf 78}, 064022 (2008)
  [arXiv:0804.4705 [hep-th]].


\bibitem{mars}
M.~Mars,
  Class.\ Quant.\ Grav.\  {\bf 16}, 2507 (1999)
  [gr-qc/9904070].



\bibitem{hidrt}
J.~Podolsky and M.~Ortaggio,
  Class.\ Quant.\ Grav.\  {\bf 23}, 5785 (2006)
  [gr-qc/0605136].

\bibitem{dias}
O.~J.~C.~Dias and H.~S.~Reall,
  Class.\ Quant.\ Grav.\  {\bf 30}, 095003 (2013)
  [arXiv:1301.7068 [gr-qc]].

\bibitem{asympflat}
M.~Ortaggio, V.~Pravda and A.~Pravdova,
  Phys.\ Rev.\ D {\bf 80}, 084041 (2009)
  [arXiv:0907.1780 [gr-qc]].

\bibitem{kundt1}
A.~Coley, R.~Milson, N.~Pelavas, V.~Pravda, A.~Pravdova and R.~Zalaletdinov,
  Phys.\ Rev.\ D {\bf 67}, 104020 (2003)
  [gr-qc/0212063].

\bibitem{kundt2}
J.~Podolsky and M.~Zofka,
  Class.\ Quant.\ Grav.\  {\bf 26}, 105008 (2009)
  [arXiv:0812.4928 [gr-qc]].

\bibitem{hyporthog}
H.~S.~Reall, A.~A.~H.~Graham and C.~P.~Turner,
  Class.\ Quant.\ Grav.\  {\bf 30}, 055004 (2013)
  [arXiv:1211.5957 [gr-qc]].


\bibitem{bianchi}
V.~Pravda, A.~Pravdova, A.~Coley and R.~Milson,
  Class.\ Quant.\ Grav.\  {\bf 21}, 2873 (2004)
  [Erratum-ibid.\  {\bf 24}, 1691 (2007)]
  [gr-qc/0401013].

\bibitem{ricci}
M.~Ortaggio, V.~Pravda and A.~Pravdova,
  Class.\ Quant.\ Grav.\  {\bf 24}, 1657 (2007)
  [gr-qc/0701150].

\bibitem{ghp}
 M.~Durkee, V.~Pravda, A.~Pravdova and H.~S.~Reall,
  Class.\ Quant.\ Grav.\  {\bf 27}, 215010 (2010)
  [arXiv:1002.4826 [gr-qc]].

\bibitem{mann03}
R.~B.~Mann and C.~Stelea,
  Class.\ Quant.\ Grav.\  {\bf 21}, 2937 (2004)
  [hep-th/0312285].

\bibitem{klr}
H.~K.~Kunduri, J.~Lucietti and H.~S.~Reall,
  Phys.\ Rev.\ D {\bf 74}, 084021 (2006)
  [hep-th/0606076].


\end{thebibliography}
\end{document}